\documentclass[twocolumn]{aastex62}

\received{}
\revised{}
\accepted{}
\submitjournal{ApJ}

\usepackage[]{natbib}
\usepackage{graphicx}
\usepackage{amssymb}
\usepackage{amsfonts}
\usepackage{scalerel}
\usepackage{tikz,xcolor,hyperref}

\definecolor{lime}{HTML}{A6CE39}
\DeclareRobustCommand{\orcidicon}{
	\begin{tikzpicture}
	\draw[lime, fill=lime] (0,0) 
	circle [radius=0.16] 
	node[white] {{\fontfamily{qag}\selectfont \tiny ID}};
	\draw[white, fill=white] (-0.0625,0.095) 
	circle [radius=0.007];
	\end{tikzpicture}
	\hspace{-2mm}
}

\foreach \x in {A, ..., Z}{\expandafter\xdef\csname orcid\x\endcsname{\noexpand\href{https://orcid.org/\csname orcidauthor\x\endcsname}
			{\noexpand\orcidicon}}
}

\newenvironment{enumerate*}%
  {\begin{enumerate}%
    \setlength{\itemsep}{0pt}%
    \setlength{\parskip}{0pt}}%
  {\end{enumerate}}

\def\integral{{\it INTEGRAL}}
\def\integrale{{\it INTEGRAL }}
\def\xte{{\it RXTE}}
\def\xtee{{\it RXTE }}
\def\gro{{\it CGRO}}

\def\swift{{\it Swift}}
\def\swifte{{\it Swift }}

\def\suzakue{{\it Suzaku }}

\def\nust{{\it NuSTAR}}
\def\nuste{{\it NuSTAR }}

\def\apjl{\rm ApJL}
\def\apjs{\rm ApJS}

\def\aap{\rm A\&A}

\shorttitle{Accretion modes of Cyg X-1}
\shortauthors{Lubi\'nski et al.}

\begin{document}

\title{Distinct accretion modes of Cygnus X-1 revealed from hard X-rays}

\correspondingauthor{Piotr Lubi\'nski}
\email{p.lubinski@if.uz.zgora.pl}

\author{Piotr Lubi\'nski\orcidA{}}\affil{Institute of Physics, University 
of 
Zielona G\'{o}ra, Licealna 9, 
PL-65-417 Zielona G\'{o}ra, Poland}

\author{Alexandros Filothodoros\orcidB{}}
\affil{Institute of Physics, University of Zielona G\'{o}ra, Licealna 9, 
PL-65-417 Zielona G\'{o}ra, Poland}
 
\author{Andrzej A. Zdziarski\orcidC{}}
\affil{Nicolaus Copernicus Astronomical Center, Polish Academy of Sciences,
Bartycka 18, PL-00-716 Warszawa, Poland}

\author{Guy Pooley}
\affil{Astrophysics Group, Cavendish Laboratory, 19. J. J. Thomson Avenue,
Cambridge CB3 0HE, UK}

\begin{abstract}

Thanks to recurrent observations of the black hole binary Cyg X-1 carried out 
over 15 years the \integrale satellite has collected the largest data set in the 
hard X-ray band for this source. We have analyzed these data, complemented by 
data collected by other X-ray satellites and radio flux at 15 GHz. To 
characterize the spectral and variability properties of the system we have 
examined parameters such as the hard X-ray flux, photon index and fractional 
variability. Our main result is that the 2D distribution of the photon index and 
flux determined for the 22--100 keV band forms six clusters. This result, 
interpreted within the Comptonization scenario as the dominant process 
responsible for the hard X-ray emission, leads to a conclusion that the hot 
plasma in Cyg X-1 takes the form of six specific geometries. The distinct 
character of each of these plasma states is reinforced by their different X-ray 
and radio variability patterns. In particular, the hardest and softest plasma 
states show no short-term flux - photon index correlation typical for the four 
other states, implying a lack of interaction between the plasma and accretion 
disk. The system evolves between these two extreme states, with the spectral 
slope regulated by a variable cooling of the plasma by the disk photons. 

\end{abstract}


\section{Introduction}
\label{intro}

Binary systems hosting a black hole (BH) can be subdivided into persistent and 
much more numerous transient sources (e.g., \citealt{Tetarenko2016}). Both 
classes are observed in several X-ray spectral states, followed by distinct
properties of the emission from radio to UV bands (e.g., \citealt{Belloni2016}).
The three basic states are the soft (or high) state (SS) with a stronger soft 
X-ray emission, the hard (or low) state (HS) when the hard X-ray emission 
dominates, and the intermediate state (IS) corresponding to a transition between 
the two other states. There is a consensus that these states are associated with 
various arrangements of the accretion disk, plasma region(s) and outflowing 
material, the last in the form of jets or winds (e.g., \citealt{Fender2016}). 
However, there are many different propositions of the actual scenario of the 
system's geometry evolution. Prevailing are those invoking a varying inner 
radius of the disk as a main driver of the system changes (e.g., 
\citealt{Esin1997}), i.e., related to the advection-dominated accretion flow 
(ADAF) models \citep{Narayan1994,Abramowicz1995} with a hot inner flow replacing 
the disk. Alternative to the truncated disk scenario are, for example, models 
with a magnetized disk of constant size where the state transitions result from 
a varying strength of the toroidal component of the disk field 
\citep{Begelman2015}. 

Studies of spectral states of BH binaries (BHB) were intensified in the 
mid-1990s when the Rossi X-ray Timing Explorer (\xte, \citealt{Bradt1993}) 
started operation. Besides a detailed modeling of the high-quality spectra there
were several phenomenological tools developed: the hardness-intensity diagrams 
(HIDs), the fractional rms variability (RMS)-intensity diagrams, and the 
hardness-RMS diagrams (HRD). Whereas transient BHBs typically follow the 
so-called q-shaped track in HID, persistent BHBs show rather only a fraction of 
that track \citep{Belloni2016}. 
 
Varying energy spectra and variability explored through the RMS, power-density 
spectra (PDS), coherence and lags studies resulted in an extension of the state 
classification. Many transient sources exhibit a phase of very strong disk 
emission coupled with a prominent steep power-law continuum at higher energies, 
so-called very high state or steep power-law state (VHS, e.g., 
\citealt{McClintock2006}). Transitions between the SS and HS states were found 
to show two different phases, introducing the soft intermediate (SIMS) and hard 
intermediate (HIMS) states instead of a single IS for transient systems (e.g., 
\citealt{Belloni2010}). 

An unambiguous, though simple way of state classification is vital to get deeper 
insights on the physics of accreting BHBs. The most reliable selection is 
usually obtained with the criteria based on color-color or slope-slope 
diagrams, whenever the X-ray spectra  are available (e.g. 
\citealt{Zdziarski2002b,Wilms2006,Gierlinski2010}). Nevertheless, to 
fully explore the evolution of both transient and persistent systems, the X-ray 
monitoring data and HRD-based selection must be applied, at the expense of a 
somewhat weaker identification of the intermediate state \citep{Grinberg2013}.

Cyg X-1, the first BHB detected \citep{Bowyer1965} and one of the brightest,
persistent hard X-ray sources, hosts a black hole and a blue supergiant HD 
226868. Thanks to the brightness of Cyg X-1, a wealth of information about its 
nature has been gathered. However, being one of only several known high-mass 
wind-fed persistent BHBs, it might be not quite representative of the BHB class 
in general. Indeed, the HID diagram for Cyg X-1 does not show a typical q-track, 
occupying only a part of the region covered typically by the transient systems 
\citep{Belloni2010}. This can be related to a relatively small range of the 
bolometric luminosity observed for this source, varying by a factor $\lesssim 
10$ \citep{Wilms2006,Zdziarski2011}. The transitions between the soft and hard 
state occur at the same flux level in both directions, whereas the transient 
systems display a hysteresis behavior. Moreover, during the soft state Cyg X-1 
does not reach very low level of rms variability, typical of that state in other 
systems \citep{Belloni2010}.

Variability studies at soft X-rays revealed that the hard state in which Cyg X-1 
is observed for most of the time changed for a quite long period into a slightly 
softer hard state, showing a number of failed transitions into the soft state 
\citep{Pottschmidt2003a}. Cyg X-1 belongs to the anomalous track class, 
exhibiting a stronger radio-soft X-ray correlation than that typically observed 
for the transient BHBs \citep{Zdziarski2011}. In the hard X-ray band this 
correlation becomes more complex, with the highest radio emission observed for 
medium X-ray fluxes \citep{Wilms2006,Zdziarski2011}.

The vast majority of extensive studies of Cyg X-1 spectral states was based on the 
\xtee data, predominantly limited to the 3--35 keV band. Emission in this band 
is commonly thought to be a mixture of the primary emission of the disk and hot 
plasma, accompanied by Compton reflection of the plasma radiation from the disk.
Therefore, an interpretation of the results in terms of a complex physical model 
is usually demanding. On the other hand, the emission in the 22--100 keV band is 
dominated by the Comptonized continuum (see fig 2 of \citealt{Filothodoros2018}), 
allowing for a more direct analysis of the plasma properties. There were many 
spectral studies of the hard X-ray band exploring high-quality data from pointed 
observations (e.g. \citealt{Gierlinski1997,DelSanto2013}) and various examples 
of a comprehensive analysis of a large data set covering a long period (e.g., 
\citealt{Gleissner2004b,Zdziarski2011}). 

\integrale satellite \citep{Winkler2003} observations provided a variety of 
valuable information on the nature of Cyg X-1 (e.g. 
\citealt{Pottschmidt2003b,Zdziarski2012b,DelSanto2013}). \integral's 
uninterrupted observation (science window) typically lasts 0.5--2 hours and the 
sensitivity of the ISGRI detector \citep{Lebrun2003} allows one to obtain, within this 
period, data of a quality surpassing the daily-averaged data from the hard X-ray 
monitors. This motivated us to perform an extensive analysis of the spectral 
states of Cyg X-1, based on the hard X-ray emission. A detailed analysis of the 
state-wise summed spectra with advanced Comptonization models must be postponed 
until the improved ISGRI calibration (OSA 11, see Sec. \ref{inteda}) is extended 
to the entire mission period.

\section{Data selection and reduction}
\label{dataset}

\subsection{INTEGRAL}
\label{inteda}

Our analysis utilizes all public \integrale data released before March 1, 2018. 
These data were collected during the spacecraft orbits (revolutions, Revs), 
ranging from 22 up to 1882, i.e., within a period between 2002-12-18 08:55:11 
UTC and 2017-11-08 03:04:31 UTC. \integrale data were reduced with the OSA 
software package \citep{Courvoisier2003}, version 10.2 (ISGRI and SPI detectors) 
and version 11.0 (JEM-X 1 detector). 

To investigate the hard X-ray properties of the spectral states of Cyg X-1 we  
used a large dataset collected by the ISGRI detector in the 22--100 keV band. A 
main tool to study the spectral states is the hardness-intensity diagram. 
However, the hardness ratios depend strongly on the choice of the two energy 
bands. On the other hand, there are substantial changes of the ISGRI 
characteristics observed over the considered 15 years \citep{Natalucci2017}. For 
this reason, we decided to use the photon index of a power-law model fitted to 
the 22--100 keV spectra instead of the hardness ratio. We have also computed the 
flux in the same band based on the fitted model. In this way, by applying a set 
of time dependent instrument responses we have reduced the influence of the 
evolution of the ISGRI detector on both our basic parameters, characterizing the 
intensity and spectral slope. 

The ancillary calibration files (ARFs) of OSA 10.2 for ISGRI were released in 
December 2015 and are not appropriate for data collected more than several 
months afterwards. To enable using data collected during the last two years of the 
analyzed period, we prepared our own spectral calibration. A set of 5 ARFs was 
generated by inverting the Crab spectra collected during the years 2016 and 
2017, to ensure an overall agreement with the INTEGRAL's standard model of the 
Crab spectrum, based on the SPI detector results \citep{Jourdain2008}. Although 
this approach can be insufficient for a detailed spectral analysis with complex 
models, it ensures a good correction of the fluxes measured in relatively wide 
energy band. 

The new OSA package, version 11, was released on October 19, 2018 and is valid 
only for the Cyg X-1 observations between Revs.~1626 and 1882. Using Crab 
observations from Revs. 1662--1887 we compared spectral fits obtained for OSA 
10.2 with our ARFs and with OSA 11. Typically the Crab photon index in the 
30--100 keV band from OSA 11 is less by 0.02--0.03 than the photon index fitted 
to the OSA 10.2 spectra. This difference is a result of the new Crab spectral 
model adopted for OSA 11\footnote{L. Natalucci, private communication.}, with 
the reference photon index of 2.1 instead of 2.08 used before 
\citep{Jourdain2008}. We observe a similar effect of an overall spectral 
softening for Cyg X-1 data taken during the OSA 11 validity period. To avoid 
mixing various calibration versions for this project we used only OSA 10.2, with 
our ARFs for the recent observations.

The selection of the ISGRI data was based on a criterion of at least 10\% of the
detector area being illuminated by Cyg X-1, which corresponds roughly to 
observations with an off-axis angle $<$ 15\degr. This resulted in 8128 science 
windows filtered with the OSA software. For further analysis we selected 7907 
science windows with a relative uncertainty of the count rate in the 22--100 keV 
band $<$ 10\%. For each of them we extracted a 6-channel spectrum (with energy 
limits: 22.1--25.0--30.2--40.3--52.7--70.4--100.1 keV), analyzed later with the 
XSPEC fitting code, version 12.9.0n. The spectra were fitted with a power-law 
model, convolved with the \texttt{cflux} model to compute the flux and its 
uncertainty. The results are the hard X-ray photon index $\Gamma_{\rm H}$ and 
the flux $F_{\rm H}$ integrated in the 22--100 keV band. After this step, in 
order to reduce the uncertainty of the points in the $\Gamma_{\rm H}$-$F_{\rm 
H}$ diagram we excluded all results with a relative $\Gamma_{\rm H}$ error $>$ 
10\%, obtaining 7844 science windows. Finally, we also excluded 23 science 
windows collected during revolutions 1554--1557 when the count rate of the 
outbursting V404 Cygni in the 22--40 keV band was higher than 1000 cps. After 
applying all the selection criteria the final data set consisted of 7821 science 
windows, with a total exposure time of 18.53 Ms (covering 3.9\% of the entire 
studied period) and about 2.1 billion photons emitted by Cyg X-1 in the 22--100 
keV band. The neglected low-quality data correspond mostly to the soft state, 
decreasing its population by about 20\%. As we have tested, an exclusion of 
those 307 science windows does not introduce any bias is our results.

Despite their relatively low sensitivity, the JEM-X detectors ensure exactly 
contemporary observations to the ISGRI detector. The JEM-X field of view is 
smaller than that of IBIS and we have used only data from observations when the 
source off-axis angle was below 5\degr. Tests done with the Crab data revealed 
a small discrepancy between the hardness ratios of both JEM-X detectors. To 
avoid artifacts when mixing the JEM-X 1 and JEM-X 2 results we decided to use 
only JEM-X 1 data. We have used the standard OSA 11.0 software settings except 
for switching off the ``skipSPEfirstScw'' option and filtered out 2992 good 
science windows. The light curves of JEM-X 1 detector were extracted in two 
energy bands: 3--5 keV and 5--12 keV.

In the case of the SPI detector we used the \texttt{spiros} software included in 
OSA 10.2, set to the SPECTRA mode. Each spectrum was extracted with 6 linear 
energy channels in the 22--100 keV band. The background was estimated using the 
GEDSAT background model. The other options of the 
\texttt{spi\_science\_analysis} routine were set to their default values and all 
known bad pointings were automatically ignored. In total, 7772 single science 
window spectra were fitted with a power-law model. The sensitivity of SPI below 
100 keV is lower than that of ISGRI and for this reason we had to exclude a much 
larger fraction of data. After a selection based on the photon index uncertainty 
$\leq$ 0.1, 2894 science windows were used for further analysis.

\subsection{Other observatories}
\label{others}

Since the spectral state classification of Cyg X-1 is usually based on analysis
of the X-ray data below $\approx$ 20 keV,  \integral's data have to be 
complemented by data taken by other X-ray observatories. In addition, to deepen
our analysis we have also used radio data from a contemporary monitoring of Cyg 
X-1.

The All-Sky Monitor (ASM, \citealt{Levine1996}) on board the \xtee satellite 
monitored Cyg X-1 in the 1.5--12 keV band. We have used the data in the three 
ASM sub-bands, 1.5--3 keV, 3--5 keV and 5--12 keV, taken from the ASM 
Archive\footnote{http://xte.mit.edu/asmlc/ASM.html} covering the period 
contemporary to \integral's observations, MJD 52626 -- MJD 55200 (the ASM data 
taken after MJD 55200 were ignored, see \citealt{Grinberg2013}). Depending on 
our purposes, we used either the orbital or daily-averaged data.

After the \xtee mission completion the main instrument used for the soft X-ray 
monitoring is the Monitor of All-sky X-ray Image (MAXI, \citealt{Matsuoka2009}) 
on board the International Space Station. MAXI provides data in three energy 
bands: 2--4 keV, 4--10 keV and 10--20 keV. We used both the orbital and 
daily-averaged data, downloaded from the mission 
archive\footnote{http://maxi.riken.jp/sugizaki/v5l/}. The daily-averaged spectra
of the MAXI's Gas Slit Camera (GSC) were extracted with the \texttt{mxproduct} 
tool included in the HEASOFT package \citep{Matsuoka2009}.

BATSE data were downloaded from the Earth Occultation Data Products 
archive\footnote{https://heasarc.gsfc.nasa.gov/docs/cgro/batse/}. The daily 
spectra were extracted following the Earth occultation analysis described in 
\citet{Harmon2004}. We have used the HEASOFT \texttt{bod2pha} tool, setting the 
minimum number of flux determinations to 4 for each module of the large-area 
detector (LAD). The spectral response for each viewing period and each LAD 
module was generated with the HEASOFT \texttt{bod2rmf} tool. In total, we got 
2728 daily spectra for a period between May 2, 1991 and September 15, 1999. The 
spectra of each LAD module were fitted together in the 22--100 keV band with a 
power-law model. The photon index for a given day was computed as a weighted 
mean of photon indices fitted to all individual module spectra.

The BAT detector \citep{Krimm2013} on board the Neil Gehrels \swifte satellite 
\citep{Gehrels2004} operates in an energy band similar to \integral's ISGRI, 
namely 14--195 keV. Its sensitivity is slightly lower than that of ISGRI and its 
mean exposure time for a single day monitoring of Cyg X-1 is 91$\pm$70 minutes,  
i.e. of the order of typical duration of a single \integrale science window. We 
used BAT's continuous monitoring data in the 15--50 keV 
band\footnote{https://swift.gsfc.nasa.gov/results/transients}
. 

The largest radio dataset of the Cyg X-1 monitoring contemporary to the 
\integrale mission is that collected by the Ryle Telescope (RT, 
\citealt{Jones1991}, before MJD 53903) and its successor, the Arcminute 
Microkelvin Imager (AMI, \citealt{Zwart2008}, after MJD 54752). The RT bandwidth 
was 0.35 GHz at 15 GHz, whereas AMI covers a frequency range between 13.5 - 18 
GHz, also centered at 15 GHz. A typical exposure time was $\approx$ 10 minutes 
and Cyg X-1 was observed $\approx$ 10 times per day. The Ryle and AMI data that 
we use here were already published by \citet{Zdziarski2017,Zdziarski2020}.

\section{Results}
\label{results}

\subsection{Long-term variability of Cyg X-1}
\label{long}

\begin{figure*}
\begin{center}
\includegraphics[width=17.5cm]{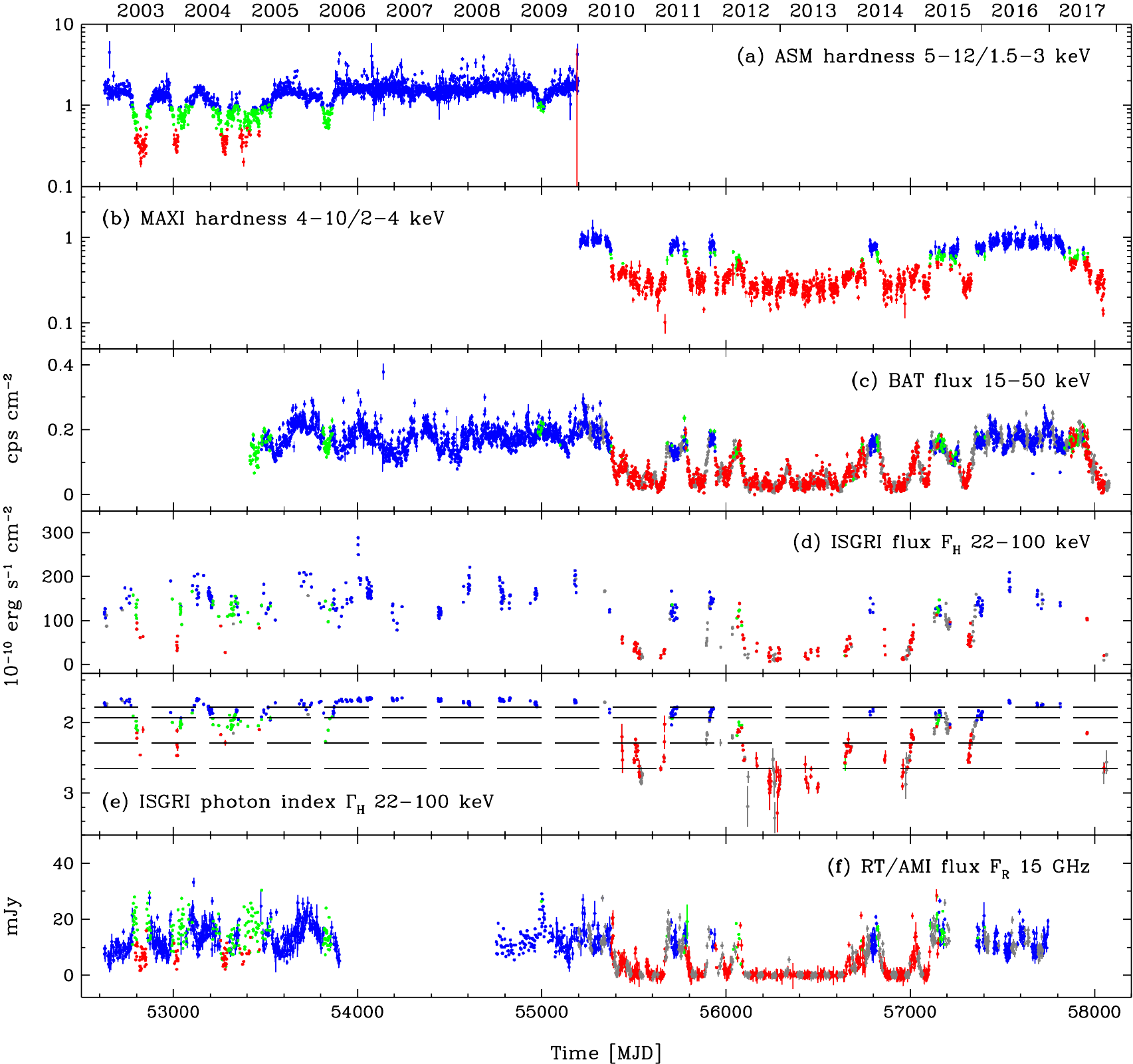}
\caption{Light curves and other spectral parameters of the Cyg X-1 emission 
obtained with various observatories during the studied period. All data are 
extracted for a time bin of one day. Color coding corresponds to the 
\citet{Grinberg2013} state classification, based on ASM data up to MJD 55200 and 
MAXI data after that day: hard (blue), intermediate (green) and soft (red). Grey 
color shows data unclassified due to the MAXI gaps. Dashed lines in panel (e) 
mark the limits of the plasma states (see Sec.~\ref{plasma}).}
\label{light}
\end{center}
\end{figure*}

In Fig.~\ref{light} we present several parameters derived from the data 
collected by the ISGRI, ASM, MAXI, and BAT detectors together with the radio 
flux measured by RT and AMI for a period of the Cyg X-1 monitoring analyzed by 
us (MJD from 52626 to 58065). All data were rebinned to a 1-day time bin. The 
points are colored according to the state selection of \citet{Grinberg2013} (see 
their table 2). Using joint ASM+MAXI data we computed that the source spent 
64\%, 10\% and 26\% of the time in the hard, intermediate and soft state, 
respectively. 

In panels (a) and (b) of Fig.~\ref{light} the ASM (5--12/1.5--3) keV hardness 
and the MAXI (4--10/2--4) keV hardness are shown, respectively, both being good
tracers of the system state. The other panels of Fig.~\ref{light} present the 
\swift/BAT flux in the 15--50 keV band (c), the ISGRI 22--100 keV flux $F_{\rm 
H}$ (d) and photon index $\Gamma_{\rm H}$ (e), and the RT/AMI 15 GHz flux (f). 
As shown in fig. 7 of \citet{Grinberg2013}, the intermediate state region 
defined for MAXI is relatively narrow and this is seen also in our results, with 
a quite narrow band of intermediate data in Fig.~\ref{light}(b). For this reason 
there are several periods after MJD 55400 during which the soft state selected 
with MAXI corresponds to a hard X-ray (BAT) and radio emission observed at 
levels typically seen during the intermediate or hard state. Taking into account 
energy bands other than the soft X-ray band, Fig.~\ref{light} demonstrates that 
the hard X-ray photon index is a very good indicator of the hard state, whereas 
the radio and hard X-ray fluxes are valuable for the soft state selection.  

\subsection{Plasma states in Cyg X-1}
\label{plasma}

\begin{figure}[]
\begin{center}
\includegraphics[height=\columnwidth,angle=270]{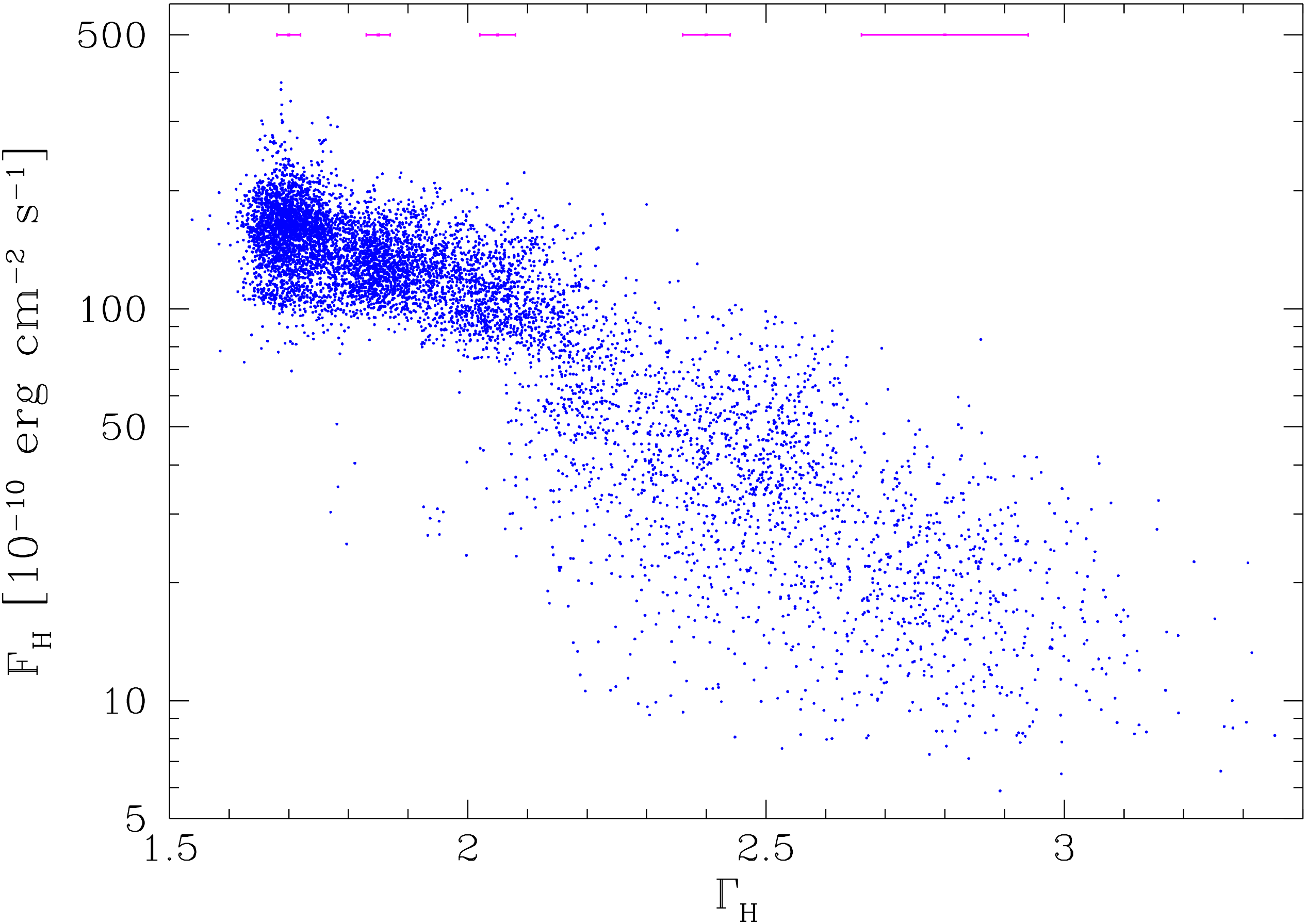}
\caption{ISGRI flux in the 22--100 keV band, $F_{\rm H}$, plotted against the 
photon index $\Gamma_{\rm H}$ characterizing the single science window spectra 
in the same energy band. Magenta lines show typical $\Gamma_{\rm H}$ errorbars 
at several $\Gamma_{\rm H}$ values (errors of $F_{\rm H}$ are very small, of the 
line width size).}
\label{gaflu}
\end{center}
\end{figure}

Figure~\ref{gaflu} presents the ISGRI 22--100 keV flux $F_{\rm H}$ as a function 
of the best-fit photon index $\Gamma_{\rm H}$ of the power-law model fitted in 
the same band to the single science window spectra. We use this plot as a hard 
X-rays counterpart of a HID used to study spectral evolution of various binary 
systems in the soft X-rays band. In general, data shown in Fig.~\ref{gaflu} form 
two distinct regions, a relatively well concentrated high-flux region 
corresponding to the hard state and a major part of the intermediate state, and 
a much more dispersed region corresponding to the soft state. 

\begin{figure*}[]
\begin{center}
\includegraphics[height=17.7cm,angle=270]{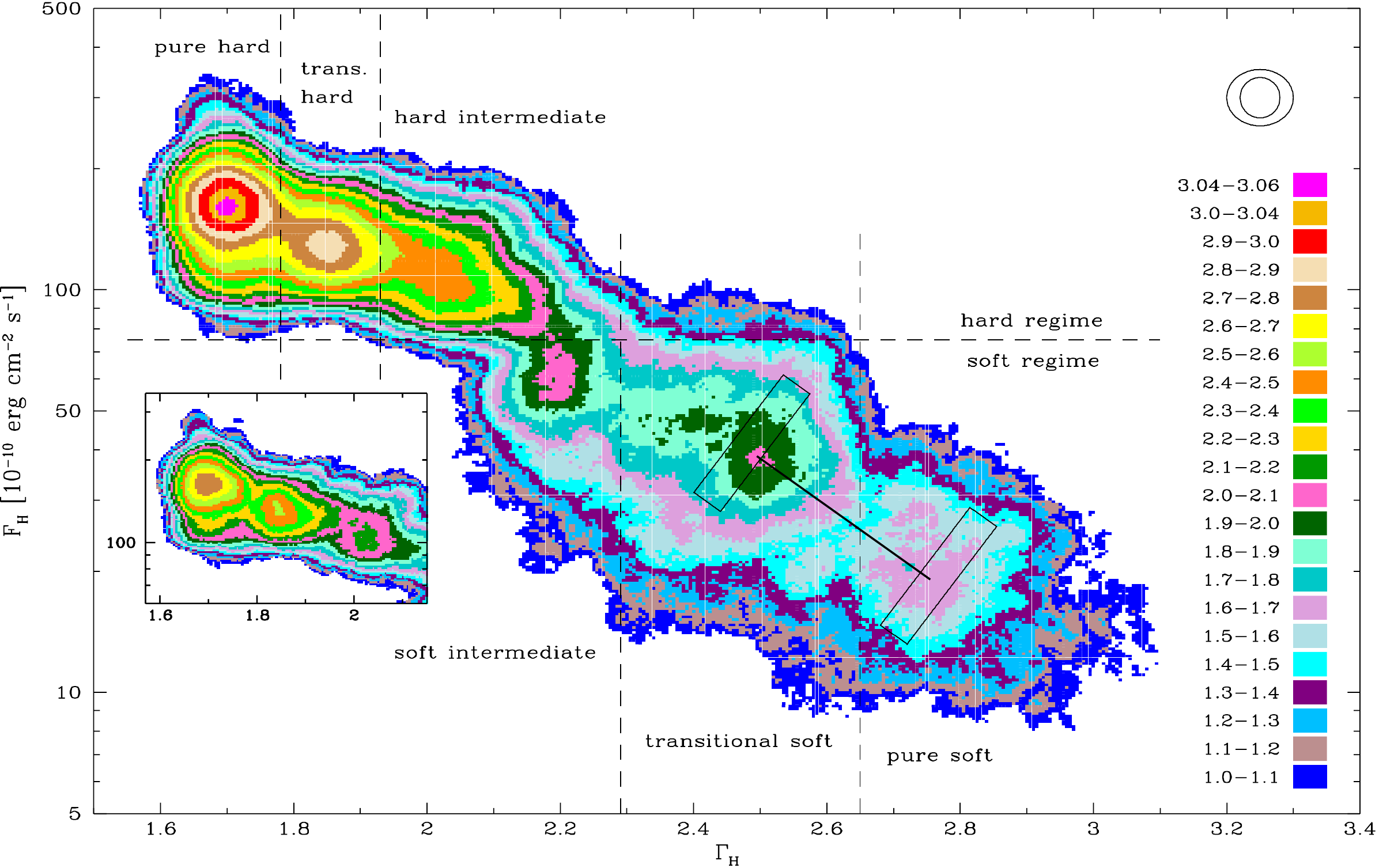}
\caption{Density map of the $\Gamma_{\rm H}$ - $F_{\rm H}$ diagram, color coding 
corresponds to the logarithm of the number of data within a region centered at 
a given pixel. A larger ellipse above the colorbar shows the size of that region. 
The vertical dashed lines mark the borders of the six plasma states, horizontal 
dashed line marks the hard/soft regime border. The solid line connecting the 
centers of the two soft state regions is shown together with two examples of the 
rectangular regions used to compute the density gradient (see the text). The 
insert shows a density map for the three hard states computed with a finer 
sampling region (smaller ellipse over the colorbar).}
\label{gfcon}
\end{center}
\end{figure*}

To better visualize the data clustering and reveal some features of the 
$\Gamma_{\rm H}$-$F_{\rm H}$ distribution we have computed its 2-D density map,
presented in Fig.~\ref{gfcon}. The density of data points was computed for 
pixels separated by $\Delta \Gamma_{\rm H}$ = 0.005 and $\Delta \log(F_{\rm H})$
= 0.007, over an ellipse with the $\Gamma_{\rm H}$ and $F_{\rm H}$ semiaxes 10 
times larger than the pixel separation. We tried several options of the pixel 
distances and integration region sizes, choosing the one best showing both local
density peaks and the regions edges or valleys between them. Also the color 
coding was adjusted to reveal those features with a rather contrasted instead of 
smoothly transiting shades.

The density map shown in Fig.~\ref{gfcon} reveals two main regions, the hard and 
soft regimes, well separated by the flux level of 75$\times 10^{-10}$ erg 
cm$^{-2}$ s$^{-1}$. In both regimes we observe three subregions, forming two 
sharp peaks for $\Gamma_{\rm H}$ $<$ 1.93, two flatter ridges in the soft-hard 
transition region (1.93 $<$ $\Gamma_{\rm H}$ $<$ 2.29) and two dispersed peaks 
for the soft data ($\Gamma_{\rm H}$ $>$ 2.29). From now we call these regions: 
pure hard (PH), transitional hard (TH), hard intermediate (HI), soft 
intermediate (SI), transitional soft (TS) and pure soft (PS) states, 
respectively, with the exact $\Gamma_{\rm H}$ and $F_{\rm H}$ limits provided in 
Table~\ref{6mean}. This nomenclature and adopted limits will be explained in 
Secs~\ref{varia} and \ref{racor}, where various properties of these states are 
discussed.

Since this is the first time when such a clustering of parameters describing the 
plasma in the accreting BH system is presented, we have to examine its 
reliability. There is a variety of simple heuristic clustering tests but most of 
them lack means to assess a relative probability between the hypotheses of 
different numbers of clusters. This limitation can be solved by using methods 
based on the probability density models with the Bayes Information Criterion 
applied to discriminate between those models \citep{Fraley2002}. In case of our 
data a construction of such density models can be demanding due to the, 
possibly, non-Gaussian shape of the regions. However, for such high density 
gradients and quite regular shapes as observed in Fig.~\ref{gfcon} those methods 
should confirm the presence of at least five clusters (except for HI state). To 
prove this we have made a simple estimate of the probability that the 
low-density valley seen between the two soft state regions appears by a chance.

The solid line in Fig.~\ref{gfcon} connects the centers of these two soft 
state regions. The valley is more or less perpendicular to this gradient line. 
We have computed the numbers of data falling into rectangular regions with 
centers located along this line. Two examples of such regions are shown in 
Fig.~\ref{gfcon}. We have tested several sizes of these rectangles. An example 
of the results of this density gradient study is presented in 
Appendix~\ref{appena}. As a null hypothesis (a single soft state) we applied 
several options of a pure or deformed Gaussian density distribution. If these 
models describe data, there are about 60 points missing in the valley (see 
Fig.~\ref{ridge}). These points must be shifted by a statistical fluctuation to 
the one of the peak regions. As shown in Fig.~\ref{gaflu}, typical 1-$\sigma$ 
uncertainty of $\Gamma_{\rm H}$ in this region of the diagram is about 0.1, 
i.e., slightly smaller than the expected shift. Thus an upper limit for a 
probability that a single point is shifted from the valley to one of the peaks 
corresponds to 1-$\sigma$ probability, 0.32. Hence the joint probability of such 
coherent fluctuation of the position of 60 points in the diagram will be smaller 
than (0.32)$^{60} \approx 2\times 10^{-30}$. 

For the other pairs of neighboring regions seen in Fig.~\ref{gfcon} the density 
gradients between them are similar or stronger in terms of the number of points 
that must be shifted (note the log-scale of the density). In addition, they have 
to be shifted by more than several $\sigma$'s (see errorbars in 
Fig.~\ref{gaflu}), thus their separation must be statistically highly probable. 
The only exception is the hard intermediate state region appearing as an 
extension of the hard state region (but see the Fig.~\ref{gfcon} insert showing
a better separation). In Secs \ref{varia} and \ref{racor} we demonstrate that 
this region exhibits several properties distinct from those of the two hard 
states.

Besides purely statistical fluctuations we considered an issue of our simplistic 
spectral model in a form of the power-law. Neglecting the continuum curvature 
due to the Compton reflection, high-energy cut-off and other effects can lead to 
an artificial photon index clustering. In \citet{Filothodoros2018} almost the 
same \integrale data set was analyzed through spectral fitting of a hybrid 
Comptonization model \texttt{eqpair}. Using their results we performed a test to 
check if the power-law model fitted to the spectra simulated with a realistic 
shape changing smoothly with the spectral hardness can induce a spurious 
aggregation of the $\Gamma_{\rm H}$ values. The test details and results are 
presented in Appendix~\ref{appenb}. Simulations demonstrated that the 
$\Gamma_{\rm H}$ distribution is modified by varying all main parameters of 
the \texttt{eqpair} model, not only those controlling the reflection and 
high-energy cut-off. Nevertheless, our test demonstrated that an approximation 
of the spectral shape with the power-law model does not introduce any specific 
$\Gamma_{\rm H}$ aggregation. Thus, our clustering results turn out to be 
essentially free from both statistical and systematic effects. 

\begin{figure}[]
\begin{center}
\includegraphics[width=\columnwidth]{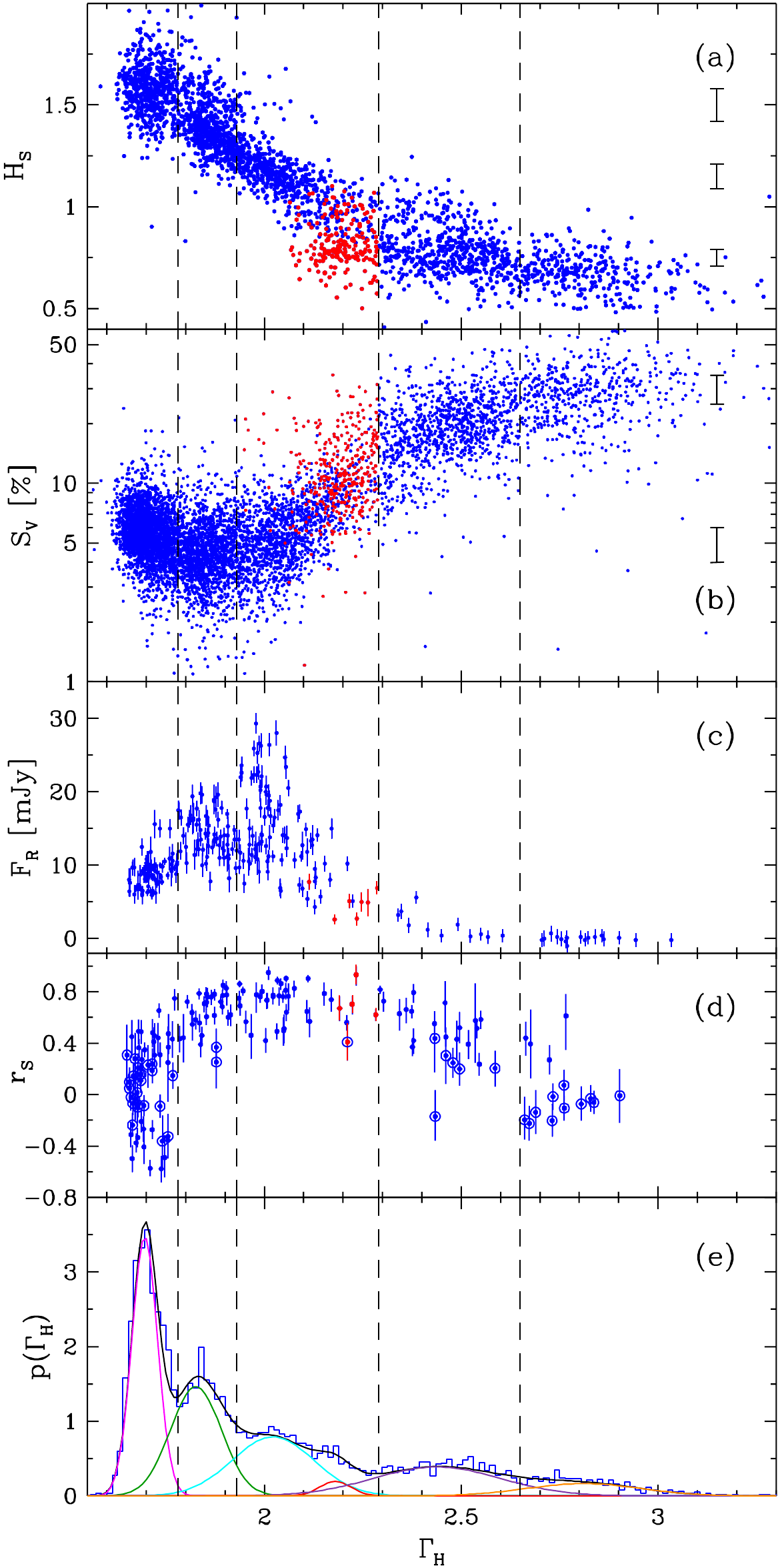}
\caption{JEM-X 1 hardness ratio $H_{S}$ (a), fractional variability amplitude 
$S_{V}$ (b), radio flux $F_{\rm R}$ (c) and the Spearman coefficient 
$r_{\rm S}$ (d) plotted against the hard X-ray photon index $\Gamma_{\rm H}$. 
The SI state data in panels (a)--(d) are shown in red. Vertical errorbars in 
panels (a) and (b) show typical errors of $H_{\rm S}$ and $S_{V}$, respectively, 
at a given level of these quantities. Circles in panel (d) mark data with the 
null hypothesis probability $>$ 0.1. (e) Decomposition of the $\Gamma_{\rm H}$ 
distribution (blue) into a sum of six Gaussians (black). The Gaussians 
corresponding to the PH (magenta), TH (green), HI (cyan), SI (red), TS (violet) 
and PS state (orange) are also shown. The dashed lines mark the $\Gamma_{\rm H}$ 
limits of these six states, except for the HI and SI states occupying the same 
range (see Table~\ref{6mean}).}
\label{hardg}
\end{center}
\end{figure}

In the rest of this section and Secs~\ref{varia} and \ref{racor} we consider 
several other parameters characterizing the Cyg X-1 emission, showing that they 
expose in various ways the distinct nature of the six states found with the 
$\Gamma_{\rm H}$-$F_{\rm H}$ density diagram. Figures \ref{hardg} and 
\ref{hardf} present these parameters as a function of $\Gamma_{\rm H}$ and 
$F_{\rm H}$, respectively. The corresponding Tables \ref{6mean} and \ref{6corr} 
summarize the results of statistical analysis of various properties of the six 
states, in terms of the mean values and correlations, respectively.

Although the density data in Fig.~\ref{gfcon} for the hard and soft regimes 
depend on both $\Gamma_{\rm H}$ and $F_{\rm H}$, the hard photon index is the 
primary parameter that enables the separation of the six states. In panel (e) of 
Fig.~\ref{hardg} we show the $\Gamma_{\rm H}$ distribution\footnote{All 
single-parameter distributions shown in this paper are computed as the 
probability density functions, i.e., taking into account the abscissa units and 
normalized to give the total integral equal to 1.} and its decomposition into 6 
Gaussians, providing a quite good approximation (reduced $\chi^{2}$ around 1.2). 
This fit should be treated with a caution because the peaks shown in 
Fig.~\ref{gfcon} are not quite symmetric and there is an overlap between both 
intermediate states. Therefore, we use the fitted curve only to guide the eye 
(see Sec.~\ref{gammas}).

The first quantity confronted with $\Gamma_{\rm H}$ and $F_{\rm H}$ is the JEM-X 
1 hardness ratio $H_{\rm S}$ (panel (a) of Figs. \ref{hardg} and \ref{hardf}). 
The JEM-X 1 data should be corrected for the orbital modulated absorption, which 
can reach quite high values \citep{Grinberg2015}. However, such a correction is 
not possible without a detailed spectral analysis, out of scope of this paper. 
When the JEM X-1 3--12 keV count rate data are plotted against the orbital phase 
(according to \citealt{Gies2008}), we do not see a clear general trend, either 
for the entire data set or state-wise selection. Therefore, we use uncorrected 
data, which introduces some scatter in the results.

As shown in Table 2, the hardness ratio $H_{\rm S}$ does not correlate with 
$\Gamma_{\rm H}$ for the PH and SI states, strongly anticorrelates for the TH, 
HI, and TS states and shows weak anticorrelation for the PS state. 
Correlations between $F_{\rm H}$ and $H_{\rm S}$ are positive and strong for 
the four softer states, very weak for TH state and weak for PH state. An 
important issue seen in Fig.~\ref{hardg}(a) is that the soft X-ray hardness 
cannot be used to identify the six plasma states found by us due to a lack of a 
clear segregation of the $H_{\rm S}$ values. In particular, the PH and TH 
states cannot be separated and the three softest states are strongly mixed. The 
two intermediate states data between $\Gamma_{\rm H}$ = 1.93 and 2.29 occupy
different ranges of the $H_{\rm S}$ range, as shown in red in 
Fig.~\ref{hardg}(a), however, this does not allow for a good separation. 

\subsection{Hard X-ray variability}
\label{varia}

\subsubsection{Fractional variability}
\label{fracvar}

The second fundamental diagram, besides the HID, commonly applied to 
characterize the behavior of BHBs is the hardness-rms diagram (e.g. 
\citealt{Belloni2010}). To construct a similar diagram with the ISGRI data, we 
have computed a normalized fractional variability amplitude $S_{\rm V}$ 
expressed as
\begin{equation}
S_{\rm V} = \big( \sigma_{\rm V}/\overline{c} \big) \times100\%,
\end{equation}
with the fractional variability amplitude $\sigma_{\rm V}$ fulfilling an equation
derived by \citet{Almaini2000}
\begin{equation}
\sum_{i=1}^{N} \frac{(c_{i}-\overline{c})^2-(\sigma_{i}^2+\sigma_{\rm V}^2)}
{(\sigma_{i}^{2}+\sigma_{\rm V}^{2})^2} = 0,
\label{eq2}
\end{equation}
where $c_{i}$ is the count rate measured for a given time bin, $\sigma_{i}$ is
its statistical error, $\overline{c}$ is the mean count rate for a given science
window, and $N$ is the number of time bins within the science window period.

The time bin adopted for this analysis was 1 minute. With this, a relative 
statistical uncertainty of the count rate for a single measurement was exceeding 
50\% only for about 3.6\% of the data. On the other hand, since the \integrale 
science window typically lasts at least 20 minutes, there were at least 20 time 
bins used to compute the $S_{\rm V}$ value for each science window. We have 
found that for 513 out of 7566 science windows the left side of Eq. \ref{eq2} was
positive even for $\sigma_{\rm V}$ = 0, i.e., the observed variability was smaller 
than purely statistical fluctuations. This indicates that the statistical error 
of the count rate determined with the OSA software is somewhat overestimated. 
For a 60 s exposure time a single ISGRI pixel registers usually 0 or 1 count 
from a source as bright as Cyg X-1. This is the lowest signal-to-noise regime, 
where the OSA software tends to sligthly amplify the statistical errors (see the 
Appendix A in \citealt{Lubinski2009} for a discussion).

To estimate this excess we have extracted 1-minute light curve for the Crab, a 
relatively stable source with the count rate level similar to that of Cyg X-1 in 
the hard state. For Crab we have found that Eq. \ref{eq2} cannot be fulfilled 
for 2380 out of 5582 science windows. Decreasing the statistical errors by a 
systematic 1\% error reduced this number to 35. We have applied the same 1\% 
correction to the Cyg X-1 data, reducing the number of non-determined 
$\sigma_{V}$ values to 212 and excluding corresponding cases from a further 
$S_{\rm V}$ analysis. The uncertainty of $S_{\rm V}$ was determined with the 
bootstrap method, in a way following that of \citet{Soldi2014}.

Figure~\ref{hardg}(b) presents the results of variability analysis where the 
$S_{\rm V}$ amplitude is plotted against the $\Gamma_{\rm H}$ index. To the best 
of our knowledge this is the first time where such diagram is constructed with 
the data from the hard X-ray band.

Our analog of the soft X-ray HRD diagram presented in Fig.~\ref{hardg}(b)
extends the findings of \citet{Grinberg2014} to energies above 15 keV. A 
complete comparison is not possible because our $S_{\rm V}$ was extracted for 
roughly 0.5--8 mHz range, whereas their fractional rms was computed for the 
0.125--256 Hz band and their $\Gamma_{1}$ was fitted below $\approx$ 10 keV. 
Nevertheless, their result for the highest energy band, 9.3--15 keV, looks
qualitatively similar to our result in a sense that the minimal variability is
observed for medium hardness spectra. Main differences are the location of that
minimum (our TH state vs. their relatively soft state) and their soft state 
variability higher than that of the hard state only for the softest data.

An analysis of Cyg X-1 \integrale data in a much broader frequency range was 
presented by \citet{Cabanac2011}, who analyzed variability in the the 27--49, 
69--90 and 96-130 keV bands with the SPI detector. Their frequency range was 1 
mHz -- 100 Hz, however, the data cover only the period from 2005 March to 2008 
May, i.e., the period when Cyg X-1 was almost exclusively in the hard state 
(see Fig.~\ref{light}). The rms amplitude found for that period increases with 
the spectral hardness (see fig.~4 of \citealt{Cabanac2011}), in agreement with 
our results shown in Fig.~\ref{hardg}(b). 

The intra-state $\Gamma_{\rm H}$-$S_{\rm V}$ correlation results presented in 
Table~\ref{6corr} confirm the trends seen in Fig.~\ref{hardg}(b), i.e., 
anticorrelation in the PH state, lack of correlation for the TH and PS states, 
and positive correlation for the three other states. For the 
$F_{\rm H}$-$S_{\rm V}$ pair the situation is different: correlation for the PH 
state, clear correlation for TH state changing into strong anticorrelation for 
the next three states and no correlation for the PS state.

\subsubsection{Short-term $\Gamma_{\rm H}$-$F_{\rm H}$ correlations}
\label{gfcor}

Inspecting the single revolution data we noticed a clear and common $\Gamma_{\rm 
H}$-$F_{\rm H}$ correlation, not seen only for the two pure states. To explore 
this topic we applied the Spearman rank-order test to the $\Gamma_{\rm 
H}$-$F_{\rm H}$ data for each revolution with at least 10 science windows, in 
total 161 revolutions. The results of this investigation revealed that a 
coherent variability lasts longer than the revolution period for a majority of 
data but sometimes there is a jump observed between the two distinct correlation 
patterns. Thus, the characteristic period of the $\Gamma_{\rm H}$-$F_{\rm H}$ 
coherence is typically longer than several days. Definitely this issue needs a 
deeper study, which is out of the scope of our paper. Nevertheless, to clean the 
correlation results from these rapid jumps for each revolution we determined the 
longest period of an interrupted correlation (or its lack) characterized by the 
minimum (or maximum) of no correlation probability, neglecting the rest of this 
revolution data. Such a selection was needed for about half of revolutions and 
neglected fraction was typically below 20\%.

\begin{figure}
\begin{center}
\includegraphics[width=\columnwidth]{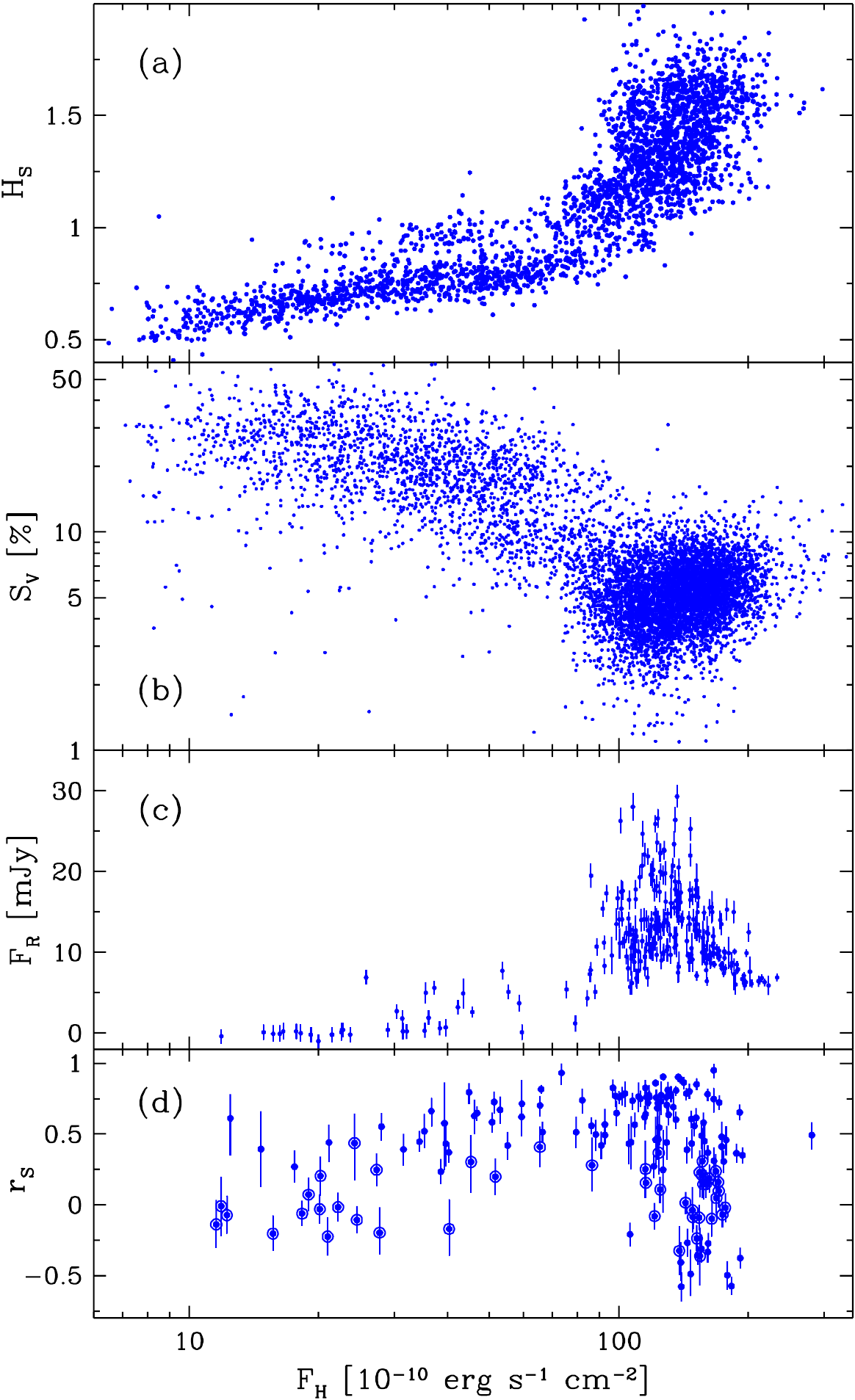}
\caption{JEM-X 1 hardness ratio $H_{S}$ (a), fractional variability amplitude 
$S_{V}$ (b), radio flux at 15 GHz $F_{\rm R}$ (c) and the Spearman coefficient 
$r_{\rm S}$ (d) plotted against the hard X-ray flux $F_{\rm H}$.}
\label{hardf}
\end{center}
\end{figure}

The results of the Spearman test are shown in Figs~\ref{hardg}(d) and 
\ref{hardf}(d), where the single revolution $\Gamma_{\rm H}$-$F_{\rm H}$ 
correlation coefficient $r_{\rm S}$ is plotted against the mean $\Gamma_{\rm H}$ 
and mean $F_{\rm H}$, respectively, for a given revolution or part of it. We do 
not show the null hypothesis probability $P_{\rm S}$ in these two figures, 
however, the points corresponding to $P_{\rm S}$ $>$ 0.1 are marked with 
circles. The errors in $r_{\rm S}$ and $P_{\rm S}$ were computed through 
Monte-Carlo simulations, taking into account the $\Gamma_{\rm H}$ and $F_{\rm 
H}$ uncertainty. Note that the $\Gamma_{\rm H}$-$r_{\rm S}$ and $F_{\rm 
H}$-$r_{\rm S}$ correlation results presented in Table~\ref{6corr} were also 
computed for the revolution-averaged data because $r_{\rm S}$ is always computed 
for a single revolution, not a single science window. On the other hand, the 
$\Gamma_{\rm H}$-$F_{\rm H}$ correlation parameters in Table~\ref{6corr} are the 
long-term results, computed for all data for a given state.

The difference between the PH state results and the TH, HI, SI and TS states 
results is striking. For the PH state a positive, missing and negative 
correlations are possible and the no-correlation probability $P_{\rm S}$ is 
quite high (see Table~\ref{6mean}). For data with the $\Gamma_{\rm H}$ between 
1.78 and 2.65 there is a dramatic change: a clear positive correlation ($r_{\rm 
S}$ $>$ 0.4) is seen and the null hypothesis probability is usually close to 0. 
For the pure soft state correlation vanishes, consistent with a trend of a 
decreasing $\Gamma_{\rm H}$-$F_{\rm H}$ coherence, observed already for the TS 
state (see also Tables \ref{6mean} and \ref{6corr}). Concluding, both the PH and 
PS states are characterized by a vanishing short-term $\Gamma_{\rm H}$-$F_{\rm 
H}$ correlation. This extends a definition that can be found in the literature 
(e.g., \citealt{Wilms2001}) identifying them with a missing (pure hard) or 
completely dominating (pure soft) disk component in the soft X-ray spectra.

\subsection{Hard X-ray - radio correlations}
\label{racor}

A large set of high quality data in the 22--100 keV band allows us to 
investigate correlations between the hard X-ray properties and the radio 
emission from Cyg X-1 in an unprecedented way. A much higher sensitivity above 
20 keV was reached only with the \gro/OSSE detector and currently with the 
\nuste satellite, however, their total exposure time spent on Cyg X-1 
observations is several orders of magnitude shorter than that of ISGRI. In
addition, since the continuous ISGRI observations are separated only by several 
minutes, there are many science windows that are truly contemporary to the radio 
observations. 

The radio emission is sizably affected by the orbital modulation effect (e.g., 
\citealt{Zdziarski2012a}). Using the ephemeris data of \citet{Gies2008} we have 
tested that the RT and AMI data show a general modulation trend when plotted 
against the orbital phase. This modulation can be well approximated by a 
sinusoid with a relative amplitude of about $\pm$15\%, very similar to that of 
\citet{Zdziarski2012a}. We have applied this correction to the radio data.

In Figures~\ref{hardg}(c) and \ref{hardf}(c) we present the 15 GHz flux $F_{\rm R}$ 
as a function of the photon index $\Gamma_{\rm H}$ and hard X-ray flux $F_{\rm 
H}$, respectively. The simultaneity criterion was the radio observation time 
being within the ISGRI observation period of a given science window. If there 
were several radio observations within that period, the radio flux was computed 
as a weighted mean. In total, there are 272 ISGRI science windows with 
contemporary radio data. We have tested that loosening the hard X-ray - radio 
simultaneity criterion from zero to several hours results in an almost unchanged 
relative data scatter, despite the sample for each state being increased by a factor 
of several. Therefore it is quite important for the X-ray/radio correlation 
study to use exactly contemporary data.

There is a limiting value of $F_{\rm H}$ seen in Fig.~\ref{hardf}(c) around 
80$\times 10^{-10}$ erg s$^{-1}$ cm$^{-2}$ separating the high and low radio 
flux levels. This value of $F_{\rm H}$ is quite close to the value separating 
the hard and soft regimes in the $\Gamma_{\rm H}$-$F_{\rm H}$ density diagram 
(Fig.~\ref{gfcon}). The highest radio fluxes are observed for medium level hard 
X-ray fluxes, whereas for the highest X-ray fluxes there is a slight 
anticorrelation between $F_{\rm H}$ and $F_{\rm R}$. A similar behavior can be 
noticed in the literature \citep{Gleissner2004b,Wilms2006,Zdziarski2011},
however, a pattern of $F_{\rm R}$ increasing and then decreasing with increasing 
$F_{\rm H}$ shown in Fig.~\ref{hardf}(c) is more clearly visible.

A more interesting correlation/anticorrelation pattern is seen in Fig. 
\ref{hardg}(c) where the radio flux is plotted against the hard X-ray photon 
index $\Gamma_{\rm H}$. General trend of the radio flux dependence on 
$\Gamma_{\rm H}$ is quite similar to that shown by \cite{Bock2011}, where the 15 
GHz flux is plotted against the soft X-ray ($\lesssim$ 10 keV) photon index. 
However, the $F_{\rm R}$ dependence on the hard X-ray photon index shows some 
stratification for the three harder states. Radio emission during the PH state 
varies in a relatively narrow, medium-flux band between 6 and 13 mJy, except for 
several points. Then, during the TH state $F_{\rm R}$ varies around about two 
times higher level. The highest $F_{\rm R}$ values are observed for the HI 
state, being on average similar to the TH state but showing the largest radio 
emission variability (see Table~\ref{6mean}). Then, the radio flux slowly 
decreases with an increasing hard X-ray photon index, vanishing above 
$\Gamma_{\rm H}$ = 2.5. A statistically significant correlation between $F_{\rm 
R}$ and $\Gamma_{\rm H}$ is found for the PH (positive) and HI and TS (both 
negative) states, as shown in Table~\ref{6corr}. A quite strong $F_{\rm 
H}$-$F_{\rm R}$ anticorrelation appears only for the PH state. 

\begin{figure}
\begin{center}
\includegraphics[width=\columnwidth]{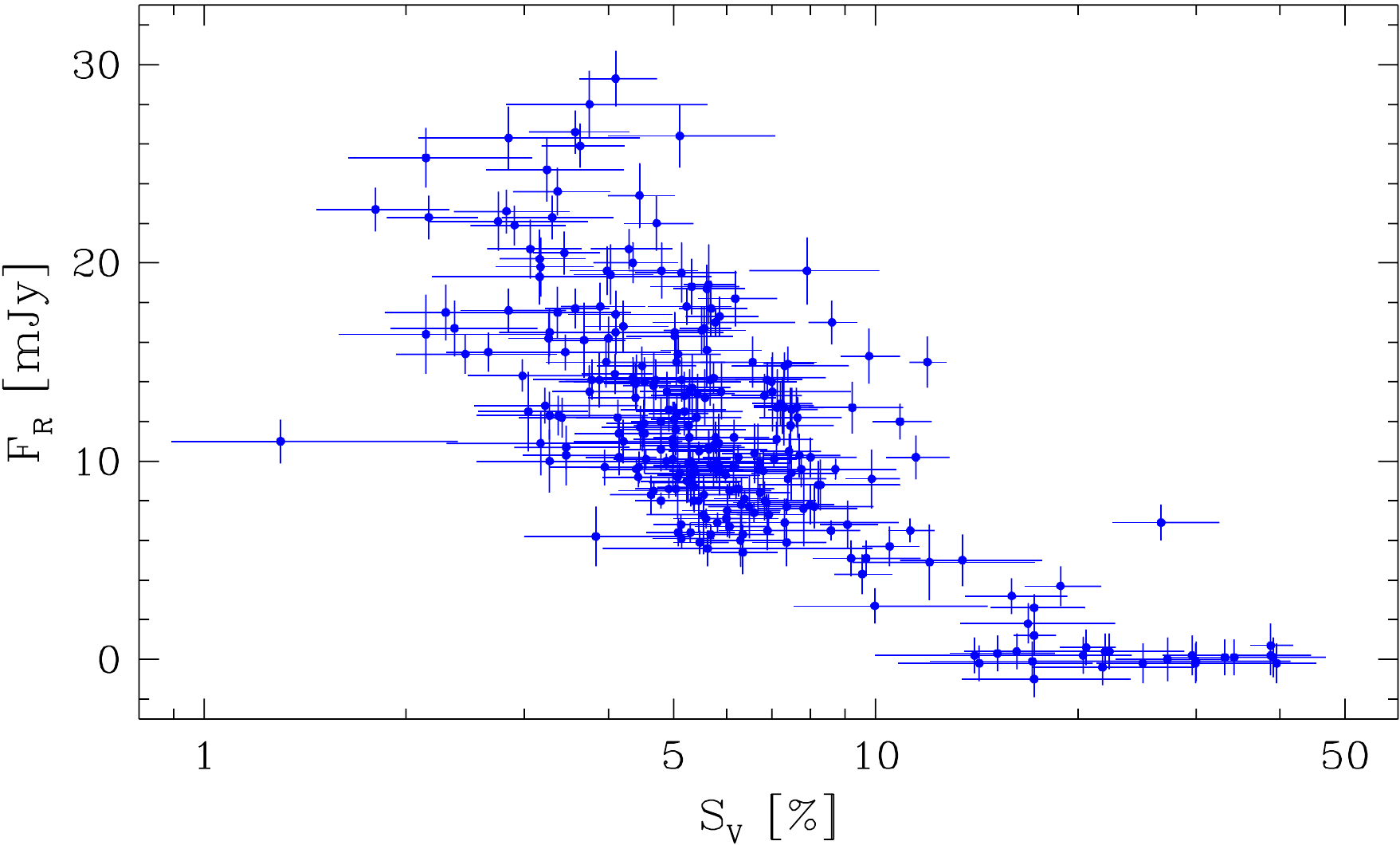}
\caption{RT/AMI radio flux at 15 GHz, $F_{\rm R}$, plotted against the fractional 
hard X-ray variability amplitude $S_{V}$.}
\label{varradio}
\end{center}
\end{figure}

The radio flux $F_{\rm R}$ plotted in Fig.~\ref{varradio} against $S_{\rm V}$ 
shows a clear overall anticorrelation, with much stronger hard X-ray variability 
at very low radio fluxes and slowly decreasing variability with increasing radio 
emission for intermediate and hard states. It is rather surprising that the 
strongest, flaring radio emission is associated with the weakest plasma 
variability. Unfortunately, we cannot investigate this issue further: although 
there were several strong radio flares observed early in the studied period (see 
Fig.~\ref{light}(g)), they were not exactly contemporary to the \integrale 
observations. When individual plasma states are considered, only the HI state 
shows a very strong $S_{\rm V}$-$F_{\rm R}$ anticorrelation (see 
Table~\ref{6corr}).

\begin{deluxetable*}{cccccccc}[t!]
\renewcommand*{\arraystretch}{0.95}
\setlength{\tabcolsep}{3pt}
\tablecaption{Selection criteria, arithmetic (first row of numbers in each 
panel) and weighted mean (second row) of parameters characterizing six plasma  
states of Cyg X-1. $\Gamma_{\rm H}$, $F_{\rm H}$, $S_{\rm V}$ - photon index,
flux and fractional variability amplitude in the 22--100 keV band, respectively,
$F_{\rm S}$ - the 3--12 keV flux, $H_{\rm S}$ - 5--12/3--5 keV hardness ratio,
$F_{\rm R}$ - flux at 15 GHz, $r_{\rm S}$, $P_{\rm S}$ - Spearman rank order 
test correlation coefficient and null hypothesis (no correlation) probability, 
respectively, for short-term $\Gamma_{\rm H}$-$F_{\rm H}$ correlations. Fraction 
of of time spent in a given state is given in \% in each panel header. Numbers 
in square brackets are per cent probabilities of transition from a given state 
to the PH, TH, HI, SI, TS, and PS state, respectively. \label{6mean}}
\tablecolumns{8}
\tablenum{1}
\tablewidth{0pt}
\tablehead{
\colhead{$\overline\Gamma_{\rm H}$} & 
\colhead{$\overline F_{\rm H}$} &
\colhead{$\overline S_{\rm V}$} & 
\colhead{$\overline F_{\rm S}$} &
\colhead{$\overline H_{\rm S}$} &
\colhead{$\overline F_{\rm R}$} &
\colhead{$\overline r_{\rm S}$} &
\colhead{$\overline P_{\rm S}$} \\
& $10^{-10}$ erg s$^{-1}$ cm$^{-2}$ & \% & counts s$^{-1}$ cm$^{-2}$ & & mJy & &
}
\startdata
\multicolumn{8}{c}{Pure hard state ($\Gamma_{\rm H}$ $\leq$1.78) 33.1\% [--,2,0,0,0,0] } \\[2pt]
1.70$\pm$0.04 & 158$\pm$33 
& 5.9$\pm$1.8 & 0.35$\pm$0.08 & 1.58$\pm$0.14 & 8.9$\pm$2.0 
& 0.07$\pm$0.34 & 0.22$\pm$0.27 \\
1.702$\pm$0.001 & 156.77$\pm$0.02 & 4.8$\pm$0.1 & 0.354$\pm$0.001 & 1.554$\pm$0.002 &
8.6$\pm$0.1 & 0.082$\pm$0.013 & 0.26$\pm$0.03 \\
\hline
\multicolumn{8}{c}{Transitional hard state (1.78 $<$ $\Gamma_{\rm H}$ $\leq$1.93) 21.5\% [3,--,4,0,0,0]} \\[2pt]
1.85$\pm$0.04 & 134$\pm$24 
& 4.7$\pm$1.8 & 0.43$\pm$0.08 & 1.40$\pm$0.14 & 14.2$\pm$2.9 
& 0.63$\pm$0.16 & 0.02$\pm$0.07 \\
1.850$\pm$0.001 & 133.60$\pm$0.03 & 3.3$\pm$0.1 & 0.416$\pm$0.001 & 1.368$\pm$0.002 &
14.0$\pm$0.1 & 0.707$\pm$0.012 &  (19$\pm$8)$\times 10^{-5}$ \\ 
\hline
\multicolumn{8}{c}{Hard intermediate state (1.93 $<$ $\Gamma_{\rm H}$ 
$\leq$2.29,
$F_{\rm H}$ $\geq$ 75$\times 10^{-10}$ erg cm$^{-2}$ s$^{-1}$) 18.7\% [0,4,--,4,1,0]} \\[2pt]
2.05$\pm$0.08 & 114$\pm$25 
& 5.9$\pm$2.4 & 0.56$\pm$0.16 & 1.14$\pm$0.14 & 15.0$\pm$6.0 
& 0.72$\pm$0.14 & 0.004$\pm$0.015 \\
2.040$\pm$0.001 & 110.98$\pm$0.03 & 4.0$\pm$0.1 & 0.565$\pm$0.001 & 1.097$\pm$0.001 &
14.1$\pm$0.1 & 0.809$\pm$0.009 & (5$\pm$2)$\times 10^{-5}$ \\ 
\hline
\multicolumn{8}{c}{Soft intermediate state (1.93 $<$ $\Gamma_{\rm H}$ 
$\leq$2.29,
$F_{\rm H}$ $<$ 75$\times 10^{-10}$ erg cm$^{-2}$ s$^{-1}$) 5.6\% [0,0,2,--,22,3]} \\[2pt]
2.18$\pm$0.07 & 49$\pm$16 & 12.2$\pm$5.5 & 0.50$\pm$0.13 & 0.82$\pm$0.12 &
5.0$\pm$1.9 & 0.67$\pm$0.19 & 0.05$\pm$0.10 \\
2.186$\pm$0.002 & 49.60$\pm$0.05 & 7.3$\pm$0.5 & 0.447$\pm$0.001 & 0.815$\pm$0.002 &
4.6$\pm$0.4 & 0.69$\pm$0.03 & 0.005$\pm$0.004 \\
\hline
\multicolumn{8}{c}{Transitional soft state (2.29 $<$ $\Gamma_{\rm H}$ $\leq$2.65) 14.1\% [0,0,1,8,--,11]} \\[2pt]
2.47$\pm$0.10 & 42$\pm$20 
& 19.8$\pm$7.2 & 0.57$\pm$0.20 & 0.79$\pm$0.12 & 1.8$\pm$1.8 
& 0.47$\pm$0.23 & 0.09$\pm$0.15 \\
2.450$\pm$0.002 & 39.49$\pm$0.03 & 12.8$\pm$0.5 & 0.487$\pm$0.001 & 0.802$\pm$0.001 &
1.7$\pm$0.3 & 0.60$\pm$0.02 & 0.06$\pm$0.02 \\ 
\hline
\multicolumn{8}{c}{Pure soft state ($\Gamma_{\rm H}$ $>$ 2.65) 7.0\% [0,0,0,2,20,--]} \\[2pt]
2.84$\pm$0.14 & 21$\pm$10 
& 28.4$\pm$9.5 & 0.52$\pm$0.15 & 0.67$\pm$0.09 & 0.0$\pm$0.4  
& 0.05$\pm$0.26 & 0.46$\pm$0.31 \\
2.797$\pm$0.005 & 19.59$\pm$0.04 & 11.5$\pm$0.8 & 0.428$\pm$0.001 & 0.661$\pm$0.002 & 
0.0$\pm$0.3 & 0.005$\pm$0.033 & 0.53$\pm$0.09 \\
\enddata
\end{deluxetable*}

\begin{deluxetable*}{ccccCcccccc}[t!]
\renewcommand*{\arraystretch}{0.95}
\setlength{\tabcolsep}{3pt}
\tablecaption{Pearson correlation coefficient (first row of numbers in each 
panel) and logarithm of the null hypothesis probability (second row) computed 
for various pairs of parameters characterizing the six plasma states of Cyg X-1.
Correlations for parameter $r_{S}$ were computed for the revolution-averaged 
$\Gamma_{\rm H}$ and $F_{\rm H}$ data. The parameter symbols are explained in 
the caption to Table~\ref{6mean}. We classify a given correlation as strong 
when its probability is above 0.99, i.e., when the logarithm of the null 
hypothesis probability is below -2.
\label{6corr}}
\tablecolumns{11}
\tablenum{2}
\tablewidth{0pt}
\tablehead{
\colhead{$\Gamma_{\rm H}$-$F_{\rm H}$} & 
\colhead{$\Gamma_{\rm H}$-$F_{\rm S}$} &
\colhead{$\Gamma_{\rm H}$-$H_{\rm S}$} & 
\colhead{$\Gamma_{\rm H}$-$S_{\rm V}$} &
\colhead{$\Gamma_{\rm H}$-$r_{\rm S}$} &
\colhead{$\Gamma_{\rm H}$-$F_{\rm R}$} &
\colhead{$F_{\rm H}$-$H_{\rm S}$} &
\colhead{$F_{\rm H}$-$S_{\rm V}$} &
\colhead{$F_{\rm H}$-$r_{\rm S}$} &
\colhead{$F_{\rm H}$-$F_{\rm R}$} &
\colhead{$S_{\rm V}$-$F_{\rm R}$}
}
\startdata
\multicolumn{11}{c}{Pure hard state} \\
$-0.09_{-0.04}^{+0.03}$ & $0.25_{-0.08}^{+0.07}$ & $-0.06_{-0.08}^{+0.10}$ & 
$-0.22_{-0.04}^{+0.04}$ & $0.14_{-0.27}^{+0.26}$ & $0.38_{-0.21}^{+0.18}$ & 
$0.11_{-0.08}^{+0.08}$ & $0.08_{-0.04}^{+0.03}$ & $0.15_{-0.27}^{+0.25}$ & 
$-0.35_{-0.19}^{+0.21}$ & $-0.23_{-0.21}^{+0.17}$ \\
-6.0 & -8.7 & -0.93 & -32 & -0.51 & -3.1 & -2.2 & -4.7 & -0.55 & 
-2.7 & -1.4 \\
\hline
\multicolumn{11}{c}{Transitional hard state} \\
$ -0.11_{-0.05}^{+0.05}$ & $0.49_{-0.06}^{+0.06}$& $-0.39_{-0.07}^{+0.06}$ & 
$0.02_{-0.05}^{+0.05}$ & $0.39_{-0.38}^{+0.28}$ & $-0.29_{-0.20}^{+0.24}$ & 
$0.01_{-0.08}^{+0.08}$ & $0.21_{-0.05}^{+0.05}$ & $0.11_{-0.39}^{+0.36}$ & 
$-0.07_{-0.24}^{+0.23}$ & $-0.17_{-0.22}^{+0.25}$ \\
-5.1 & -38 & -23 & -0.4 & -1.4 & -1.8 & -0.09 & -6.7 & -0.23 & 
-0.27 & -0.74 \\
\hline
\multicolumn{11}{c}{Hard intermediate state} \\
$-0.30_{-0.05}^{+0.05}$ & $0.49_{-0.06}^{+0.06}$ & $-0.65_{-0.04}^{+0.05}$ & 
$0.23_{-0.06}^{+0.05}$ & -$0.05_{-0.34}^{+0.36}$ & $-0.41_{-0.16}^{+0.19}$ & 
$0.33_{-0.08}^{+0.07}$ & $-0.13_{-0.06}^{+0.06}$ & $0.64_{-0.27}^{+0.17}$ & 
$0.25_{-0.20}^{+0.18}$ & $-0.63_{-0.11}^{+0.14}$ \\
-30 & -37 & -99 & -14.3 & -0.09 & -4.1 & -15 & -5.0 & -4.1 & 
-1.8 & -10 \\
\multicolumn{11}{c}{Soft intermediate state} \\
$-0.04_{-0.10}^{+0.10}$ & $0.20_{-0.14}^{+0.13}$ & $-0.07_{-0.14}^{+0.14}$ & 
$0.24_{-0.11}^{+0.11}$ & $0.15_{-1.10}^{+0.82}$ & $-0.15_{-0.71}^{+0.90}$ & 
$ 0.50_{-0.12}^{+0.09}$ & $-0.53_{-0.08}^{+0.09}$ & $0.6_{-1.4}^{+0.4}$ &
$0.15_{-0.90}^{+0.70}$ & $-0.3_{-1.0}^{+0.6}$ \\
-0.35 & -2.3 & -0.48 & -4.5 & -0.07 & -0.11 & -13 & -22 & -0.44 & 
-0.11 & -0.24 \\
\hline
\multicolumn{11}{c}{Transitional soft state} \\
$-0.17_{-0.06}^{+0.06}$ & $0.08_{-0.08}^{+0.08}$ & $-0.29_{-0.08}^{+0.08}$ & 
$0.34_{-0.06}^{+0.06}$ & $-0.44_{-0.28}^{+0.40}$ & $ -0.82_{-0.14}^{+0.43}$ & 
$0.52_{-0.07}^{+0.07}$ & $-0.30_{-0.06}^{+0.05}$ & $0.41_{-0.45}^{+0.29}$ &
$0.43_{-0.70}^{+0.40}$ & $-0.23_{-0.45}^{+0.74}$ \\
-7.3 & -1.3 & -12 & -28 & -1.5 & -2.4 & -39 & -22 & -1.3 & 
-0.67 & -0.26 \\ 
\hline
\multicolumn{11}{c}{Pure soft state} \\
$-0.24_{-0.07}^{+0.08}$ & $-0.12_{-0.11}^{+0.12}$ & $-0.18_{-0.10}^{+0.12}$ & 
$0.04_{-0.09}^{+0.09}$ & $-0.11_{-0.50}^{+0.55}$ & $0.10_{-0.53}^{+0.48}$ & 
$0.67_{-0.07}^{+0.06}$ & $0.10_{-0.06}^{+0.09}$ & $-0.27_{-0.43}^{+0.58}$ & 
$0.27_{-0.55}^{+0.42}$ & $0.30_{-0.55}^{+0.40}$ \\
-7.9 & -1.5 & -2.7 & -0.43 & -0.16 & -0.15 & -40 & -1.7 & -0.45 & 
-0.48 & -0.55 \\
\enddata
\end{deluxetable*}

\subsection{States credibility and stability}
\label{trans}

The six plasma states exposed by the $\Gamma_{\rm H}$-$F_{\rm H}$ density 
diagram in Fig.~\ref{gfcon} are separated with a high confidence degree, as 
demonstrated in Sec.~\ref{plasma}, except for the HI state. Dependence of the 
four other parameters on $\Gamma_{\rm H}$ presented in Fig.~\ref{hardg} shows 
mostly a smooth transition between the neighboring states, except for the PH and 
PS states. To reveal the differences between the states in a quantitative way we 
determined the average values of various parameters for each state, listed in 
Table~\ref{6mean}. The weighted mean values confirm a distinct character of each 
state, in particular confirming strongly a peculiarity of both pure states. The 
most convincing are the correlation parameters presented in Table~\ref{6corr}, 
revealing with a high statistical credibility a distinct dynamical character of 
each state. For example, $\Gamma_{\rm H}$ and $S_{\rm V}$ are very strongly 
anticorrelated for the PH state, for the TH state these two parameters are 
completely independent, and for the HI state a strong positive correlation 
appears. Table~\ref{6corr} provides several cases of a compelling evidence for a 
changing correlation patterns for each pair of the two adjacent states. 

Besides the physical differences of the six plasma states we examined also their 
repetitiveness. As shown in Fig.~\ref{light}(e), between MJD 52626 (December 
2002) and MJD 55342 (May 2010) Cyg X-1 was mostly in the PH state and then it 
returned to the same state for MJD 57536--57814 (May 2016 -- March 2017). The 
stability of PS state is less evident due to less abundant and more scattered 
data. Its broad peak in Fig.~\ref{gfcon} is formed mainly by data from MJD 
55527--55547 (December 2010) but Cyg X-1 was observed in this state also during 
other periods, mostly in 2013 and in November 2017. The system returns also to 
the other plasma states after long breaks. Only the SI state was observed by 
\integrale during a relatively short period of 4 years, appearing at the end of 
years 2011, 2013, 2014, and 2015, i.e., during major transitions between the 
hard and soft states.

Transitions between the six plasma states are rather rare and rapid because, as 
we tested, the single \integrale orbit data usually occupy a single state 
region. To explore this point on an hour-scale period we have computed the 
numbers of state transitions, assuming a minimal $\Gamma_{\rm H}$ change of 
0.05, i.e., larger than typical uncertainty of majority of our data. The 
transition occurence percentages for each plasma state are presented in 
Table~\ref{6mean}. There were 816 transitions found out of 7820 science windows, 
i.e., the overall state transition probability was around 0.1. The transition 
probability increases with the spectral softness, however, for the two softest states 
the $\Gamma_{\rm H}$ uncertainty is larger, potentially causing a false 
detection. The most characteristic feature of the state transition is that they 
appear almost always between adjacent states. The pure hard state shows the 
highest stability (2\% transition probability) and the smallest $\Gamma_{\rm 
H}$ changes, whereas the all three soft states are quite unstable (27, 20, and 
22\% transition probability, respectively). These results agree with the
conclusion about the stability of Cyg X-1 spectral states monitored by ASM and 
MAXI \citep{Grinberg2013}.

Another issue tested for the state transition was the flux level at which they
occur, depending on the passage direction. For the transient BH systems the
q-track is usually observed, with the hard-soft transition occuring at much 
higher flux levels than the soft-hard transition. In our data we have not found
evidence for such behavior: differences in the mean $F_{\rm H}$ flux for both
transition directions are much smaller than the standard deviation of that mean 
for each pair of the states.

\subsection{Hard X-ray photon index distribution}
\label{gammas}

At the end of this section we explore whether the finding of clustered photon index 
values could be obtained with other hard X-ray missions. There are several other 
past or current hard X-ray detectors that can be used for such a study. Among 
the two CGRO detectors, OSSE and BATSE, only the latter provided a large set of 
data in a similar energy band. However, its occultation data are of much lower 
quality than those of ISGRI. The Beppo-SAX mission has provided a rather limited 
number of pointed observations of Cyg X-1. The same is true for the HXD detector 
of Suzaku and for the most sensitive hard X-ray detectors ever launched, namely 
those onboard of \nust. On the other hand, the Swift BAT detector, characterized 
by a sensitivity similar to ISGRI, observes Cyg X-1 almost every day but its 
data are affected by a relatively large statistical uncertainty due to the 
effectively short observing period for a single day. The most promising seems to 
be an analysis of the \xtee PCA and HEXTE detectors spectra, however it is a 
rather demanding task, beyond the scope of this paper.

\begin{figure}
\begin{center}
\includegraphics[width=8.5cm]{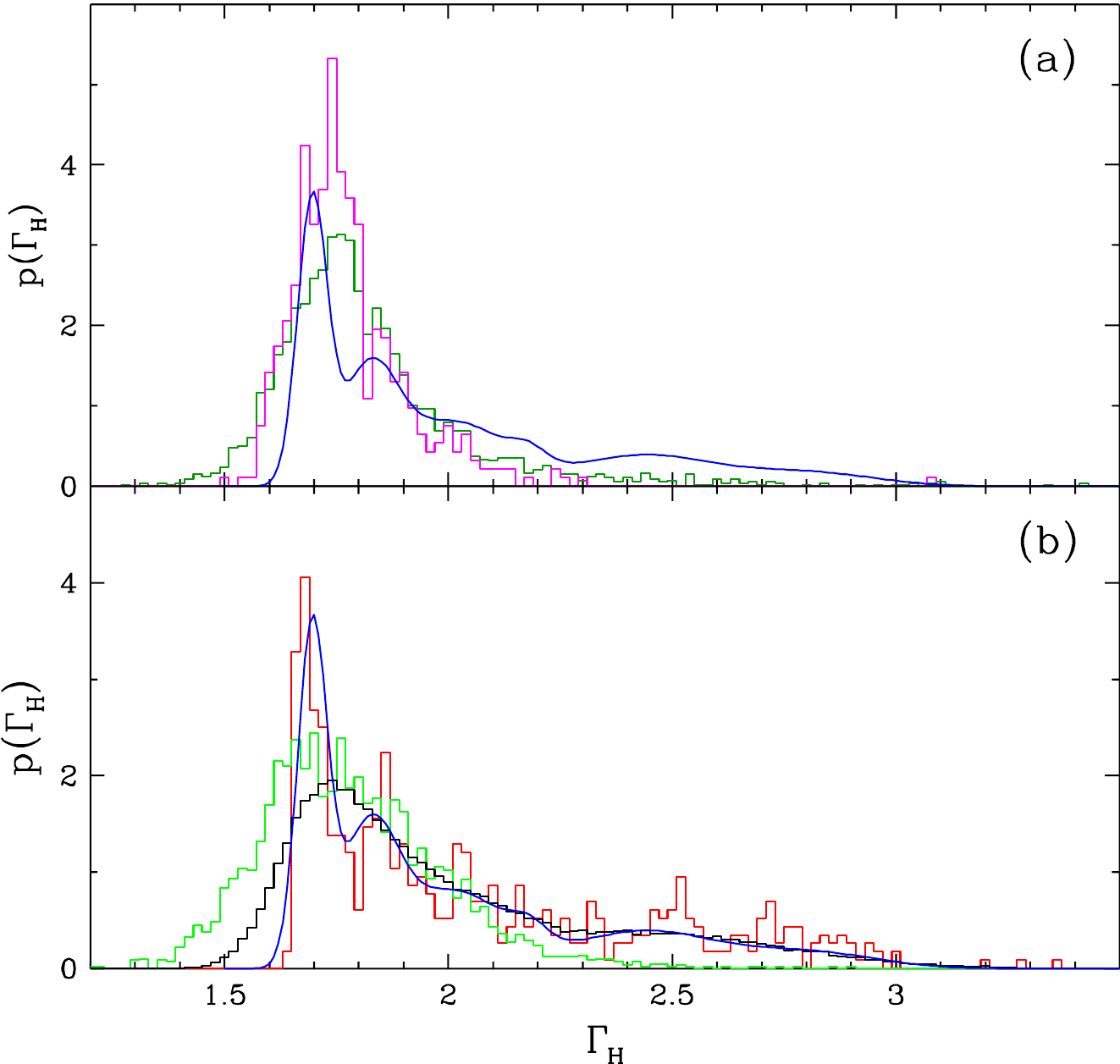}
\caption{Distributions of the 22--100 keV $\Gamma_{\rm H}$ parameter obtained 
with several alternative approaches compared with the six-Gaussian profile (blue 
curve in each panel). (a) Total (dark green) and selected (magenta, error $<$ 
0.06) photon index fitted to the BATSE spectra. (b) Photon index fitted to the 
SPI spectra (green), photon index fitted to the daily ISGRI spectra (red) and 
the ISGRI distribution blurred by Gaussian with a standard deviation equal to 
0.08 (black).}
\label{histos}
\end{center}
\end{figure}

Nevertheless, we have chosen the BATSE detector to get some hint about how its 
results for Cyg X-1 compare to those of ISGRI. Using the BATSE data has allowed 
us to extend the observing period by 9 additional years, namely 1991--2000. The 
2703 BATSE daily spectra for that period were analyzed with the power-law model. 
The uncertainty in the BATSE photon index is almost always larger than 0.05 
whereas the corresponding uncertainty in the ISGRI data is almost always smaller 
than this value. This explains why in our test we have obtained a distribution 
of the BATSE photon index resembling a broad, single Gaussian with an additional 
tail corresponding to the intermediate and soft states, shown in 
Fig.~\ref{histos}(a). We have also checked that selecting subsamples of the 
BATSE results with lower $\Delta \Gamma_{\rm H}$ values still produces 
a distribution without any evident clustering.  

The comparison between ISGRI and BATSE is also affected by an issue of the need 
to sum up the BATSE occultation data for a single day to improve the 
quality of the spectra. Any significant intraday variability of the spectral 
shape can affect the $\Gamma_{\rm H}$ distribution, washing out its discrete
features. We have tested the robustness of our results against this concern 
repeating the ISGRI spectral modeling for 579 daily-summed spectra. As 
demonstrated in Fig.~\ref{histos}(b), the positions of all states' peaks 
remained almost unchanged, when compared with the 2D distribution shown in 
Fig.~\ref{gfcon}.

Since the \integrale satellite hosts another hard X-ray instrument on its board, 
the SPI imager, it is possible to test the $\Gamma_{\rm H}$ clustering with the 
data taken at the same time as ISGRI data. We analyzed the SPI spectra with the 
power-law model in the 22--100 keV band. The number of the SPI spectra is much 
smaller than the number of the ISGRI spectra (due to an exlusion of data with 
the photon index uncertainty $>$ 0.1, see Sec.~\ref{inteda}), mostly eliminating 
the softer states' data. The SPI sensitivity below 100 keV is lower than that of 
ISGRI, resulting in an error of $\Gamma_{\rm H}$ typically around 4 times larger 
than the typical ISGRI uncertainty. Such a difference in the quality of the 
spectral data resulted in a broadened $\Gamma_{\rm H}$ distribution, see 
Fig.~\ref{histos}(b), quite similar to that produced with the BATSE data. To 
check the effect of the smearing of the distribution due to a large $\Gamma_{\rm 
H}$ uncertainty, we applied a Gaussian blur with a standard deviation of 0.08 to 
the ISGRI data. The resulting distribution is presented with a black histogram 
in Fig.~\ref{histos}(b), clearly showing that a high-precision determination of 
the photon index is crucial to uncovering its clustering for Cyg X-1.

\section{Discussion}
\label{discuss}

\subsection{Spectral and plasma states of Cyg X-1}
\label{states}

\begin{figure}
\begin{center}
\includegraphics[width=8.5cm]{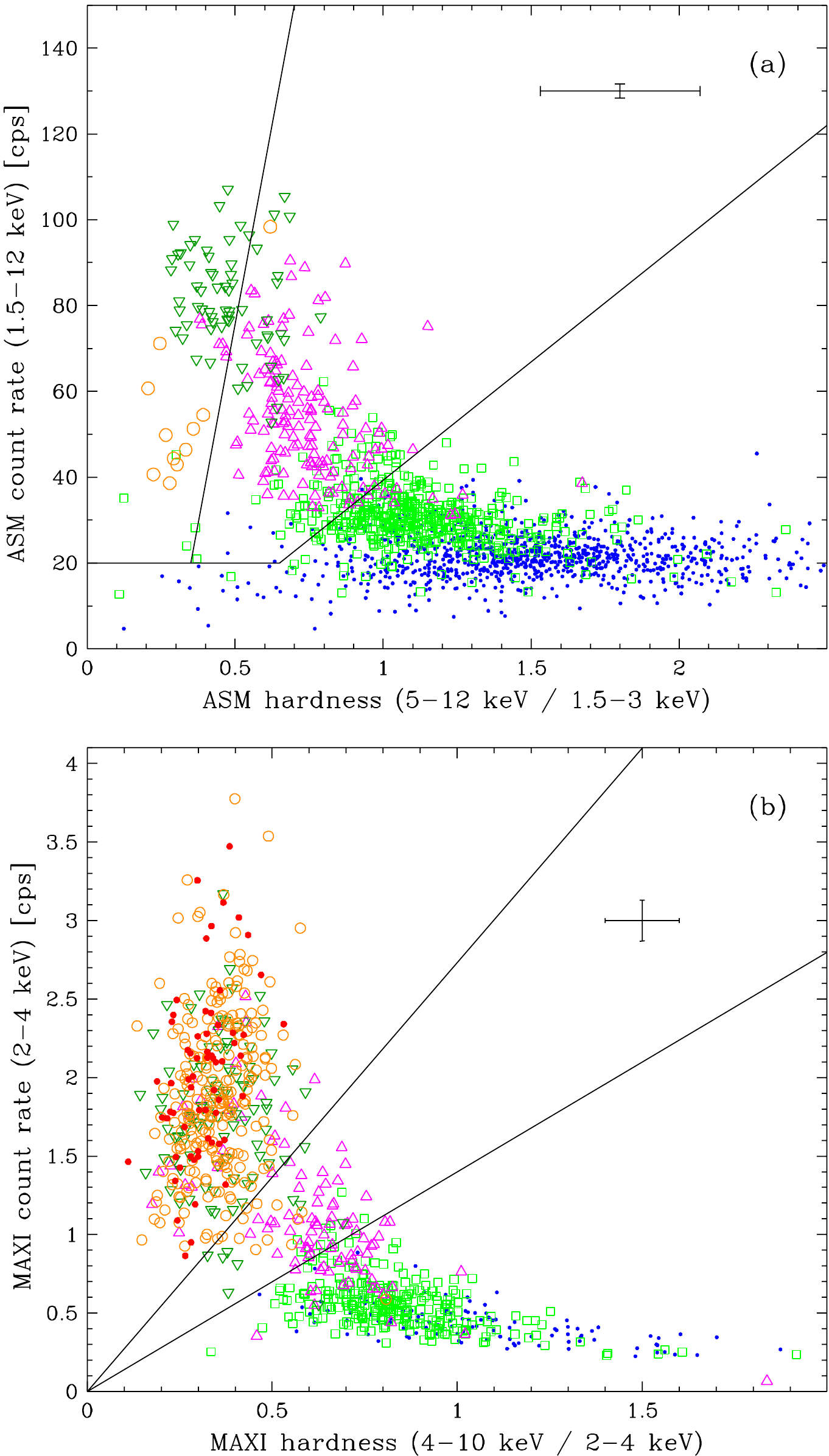}
\caption{Comparison of the state selection based on the soft and hard X-ray 
criteria. The ASM (a) and MAXI (b) data contemporary to the ISGRI data are 
marked according to the six plasma states selected with the hard X-ray photon 
index $\Gamma_{\rm H}$ criterion: PH (blue dots), TH (light green squares), HI 
(magenta triangles up), SI (dark green triangles down), TS (orange circles), 
and PS (red dots, not observed for ASM). Error bars in each panel show typical 
errors, the solid lines show the borders of the soft, intermediate and hard 
spectral state regions according to the \citet{Grinberg2013} state selection 
criteria.}
\label{boxes}
\end{center}
\end{figure}

A classification of the plasma states based on $\Gamma_{\rm H}$ should be 
confronted with a conventional spectral state selection based on the soft 
X-rays. In Fig.~\ref{boxes} we present a comparison of our state selection with 
the state definitions of \citet{Grinberg2013} derived with the ASM and MAXI  
monitors. Since the energy band and quality of monitoring data are both limited, 
when compared to the spectra from the soft X-ray telescopes, the state 
classification of \citet{Grinberg2013} was based on a comparison with the 
results of contemporary \xte/PCA observations. For ASM this gave a good 
selection of the soft and hard states, whereas the intermediate state was 
contamined up to 10\%. Selection criteria applied to the MAXI data were more 
conservative as shown in Fig.~\ref{boxes} where the intermediate state region 
for MAXI is narrower than the corresponding region for ASM.

We explored the numbers of the soft, intermediate and hard state selections 
according to the \citet{Grinberg2013} criteria for ASM and MAXI data 
contemporary to \integrale data and classified by a range of the hard X-ray 
photon index, i.e., data shown with colors in Fig.~\ref{boxes}. We observe 
similar trends as those found by \citet{Grinberg2013}, when comparing the all 
sky monitors state selection with that based on spectral fitting. The PH state 
data are practically always classified as the hard state. The same happens for 
the TH state for the MAXI-based selection, whereas for the ASM data we found 
about 20\% data classified as the intermediate state. The HI state is mostly 
also intermediate with the ASM-based selection but a majority of the HI data 
fall outside the narrow region of the intermediate state defined with the MAXI 
criteria. The SI state data are mostly found in the soft state region, 67\% for 
ASM and 90\% for MAXI. Both TS and PS states data are almost always classified 
as the soft state with the ASM and MAXI criteria.

\begin{figure}
\begin{center}
\includegraphics[width=8.5cm]{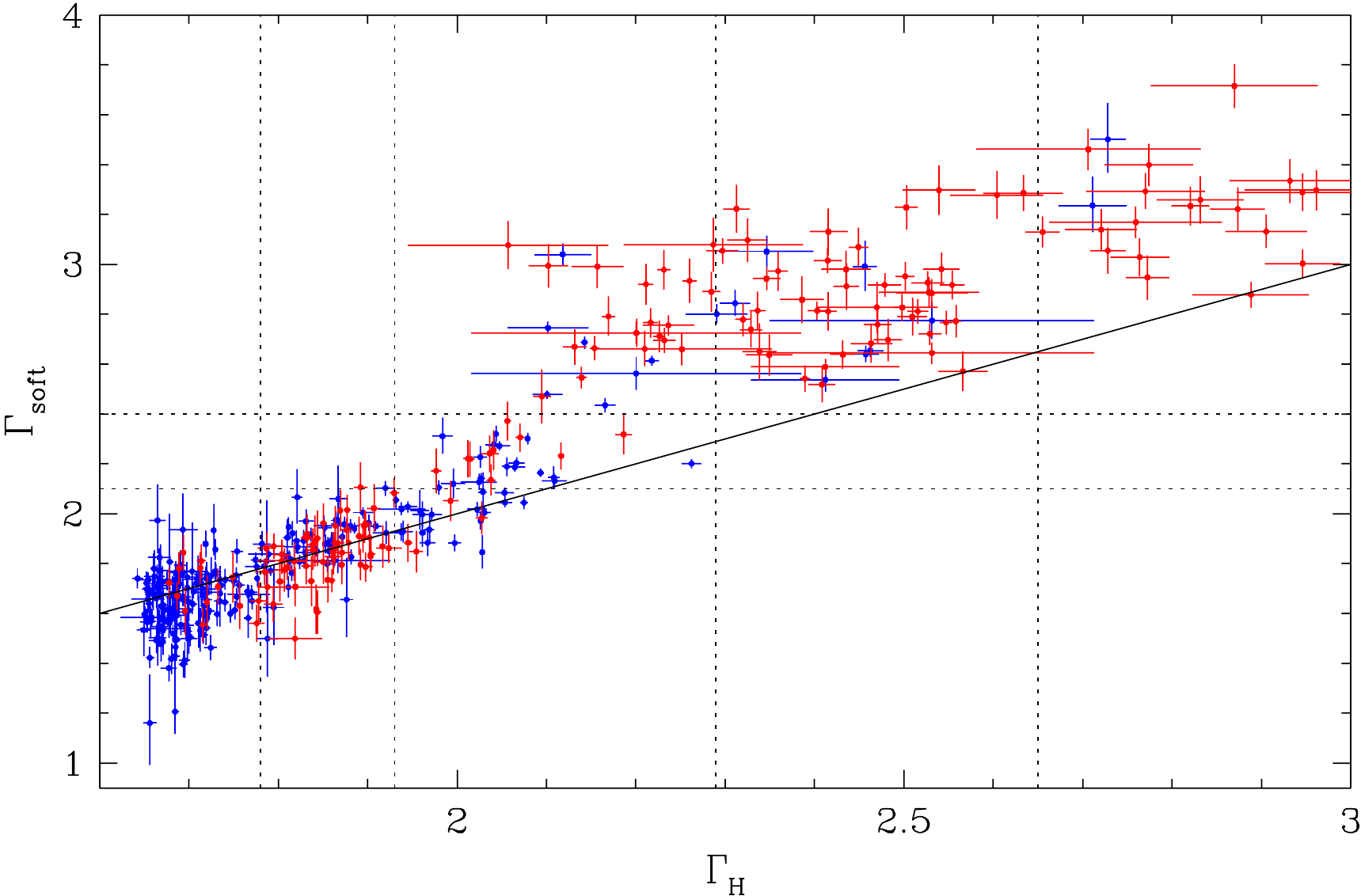}
\caption{Photon index $\Gamma_{\rm soft}$ in the 3--12 keV band computed for
the ASM (blue) and MAXI (red) daily spectra contemporary to the ISGRI data as a
function of the hard X-ray photon index. Vertical dashed lines show the borders
of six plasma states found with ISGRI. Horizontal dashed lines show the borders
of the hard, intermediate and soft states according to classification of 
\citet{Zdziarski2002b}. The solid line is the equality line.}
\label{gams}
\end{center}
\end{figure}

As mentioned in Sec.~\ref{intro}, the most reliable state classification is
obtained when the spectral slope can be determined directly from the spectra or
from monitoring data converted into fluxes in several bands. The second approach 
was developed e.g., by \citet{Zdziarski2002b}, where the ASM count rates were 
converted into flux using a redistribution matrix obtained through a comparison 
with contemporary PCA spectra. We applied the same procedure to the ASM daily 
averaged data, using the same matrix and determining fluxes in the three energy 
bands: 1.5--3.0 keV, 3.0--5.0 keV and 5.0--12.0 keV. With the last two values 
we determined the photon index in the 3.0--12.0 keV band, $\Gamma_{\rm soft}$. 
In the case of MAXI we fitted the powerlaw model to the GSC daily spectra 
obtaining directly the photon index in the same energy band. The $\Gamma_{\rm 
soft}$ results contemporary to the \integrale data are shown in Fig.~\ref{gams}, 
as a function of the hard X-ray photon index $\Gamma_{\rm H}$.

The solid line in Fig.~\ref{gams} shows that for the PH and TH states there is 
an agreement between $\Gamma_{\rm H}$ and $\Gamma_{\rm soft}$. Data are 
scattered, at least partly due to the orbital modulated absorption of the soft 
X-ray emission. For the rest of data the $\Gamma_{\rm soft}$ values are larger 
than the $\Gamma_{\rm H}$ values, what can be explained by a disk emission 
component modulating the soft X-ray part of the spectra.

The relation between the soft and hard X-ray photon indices for Cyg X-1 was 
studied by \citet{Wilms2006} who analyzed the PCA+HEXTE spectra with an absorbed 
broken power-law model, with the break energy around 10 keV and the $\Gamma_{1}$ 
and $\Gamma_{2}$ indices fitted below and above that energy, respectively. Their 
model included also a $K_{\alpha}$ iron line and an exponential cut-off of the 
high-energy power-law component. Their results are different from the 
$\Gamma_{\rm soft}$-$\Gamma_{\rm H}$ relation shown in Fig.~\ref{gams}, with 
data lying approximately along a single straight line and $\Gamma_{2}$ being 
always smaller than $\Gamma_{1}$, an effect that was ascribed to a missing 
reflection component in their model. The difference with respect to our results 
can be explained by the fact that our $\Gamma_{\rm H}$ corresponds to effectively 
larger energies than the $\Gamma_{2}$ fitted by \citet{Wilms2006} because
the combined PCA+HEXTE spectra are statistically dominated during the fit by 
high-quality PCA data below 30 keV. Thus, $\Gamma_{2}$ is probing the reflection 
peak range whereas $\Gamma_{\rm H}$, fitted above 22 keV, is less affected by 
the reflection component. The soft X-ray photon index $\Gamma_{1}$ of 
\citet{Wilms2006} and \citet{Grinberg2013,Grinberg2014} (who, in addition, 
included the disk component in their model) also cannot be directly compared 
with our $\Gamma_{\rm soft}$ index. Interestingly, the range of their 
$\Gamma_{1}$ values is practically the same as for our $\Gamma_{\rm soft}$, 
namely between 1.6 and 3.4. The 3--12 keV band is almost not affected by 
absorption and for the hard state the disk, reflection and iron line components 
are weak. Thus, for the hardest data $\Gamma_{1}$
and $\Gamma_{\rm soft}$ should be very similar.
 
Spectral state classification based on the soft X-ray results, either count 
rates or fluxes, cannot reveal the six states found with the $\Gamma_{\rm H}$ 
photon index. Figs \ref{boxes} and \ref{gams} demonstrate that the PS, TS and SI 
states occupy similar range of the soft X-ray parameters. The same holds for the 
three harder states except for the HI state, roughly traceable through the 
$\Gamma_{\rm soft}$ values. In fact, the definitions of the intermediate state 
based on the soft X-ray photon index adopted by \citet{Zdziarski2002b} (see the 
horizontal lines in Fig.~\ref{gams}) and \citet{Grinberg2013} (2.0 $<$ 
$\Gamma_{1}$ $<$ 2.5) are consistent only with our HI state. 

\citet{Pottschmidt2003a} found that in 1998 May the source switched from a 
typical hard state observed earlier with \xtee to a slightly softer hard state. 
This new hard state exhibited different variability patterns, with a smaller rms 
amplitude and relatively frequent, failed or successful transitions to the 
intermediate or soft state. Since the \citet{Pottschmidt2003a} observations were 
done before the \integrale launch, we cannot verify if their bimodal character 
of the hard state corresponds to our PH and TH states.

An energy-resolved analysis of Cyg X-1 variability with the \xtee data taken
over the 1999--2011 period \citep{Grinberg2014} partly covers the period analyzed
by \citet{Pottschmidt2003a}. This new study confirmed a distinct character of
the variability (rms, time lags) observed for the hardest spectra, with $\Gamma_{1}$ 
$<$ 1.75. As mentioned above, for the hardest data $\Gamma_{1}$ and $\Gamma_{\rm 
soft}$ should be very similar, justifying the identification of the hardest data 
of \citet{Grinberg2014} with the pure hard state. A much smaller set of Cyg X-1 
\xtee data was also analyzed in terms of a cross correlation function (CCF) 
between the photon index and the 3--20 keV count rate at 100 ms time resolution 
\citep{Skipper2013}. They found that there is an evolution of the CCF from a 
correlation seen for the hardest spectra changing into an anticorrelation for 
the rest of the hard state spectra.

In Sec.~\ref{intro} we mentioned that Cyg X-1 varies in a quite limited region
of the HID diagram when compared to the systems undergoing transitions. Our 
finding is that also in the hard X-ray diagram we do not see a hysteresis: 
transitions between the states in both directions happen at the same flux level. 
However, similarly to the transients we found two distinct intermediate states, 
hard and soft. To the best of our knowledge, bimodality of the intermediate state is 
almost not explored for Cyg X-1 because in the soft X-rays the HI state just 
forms the canonical intermediate state, whereas the SI state falls into the soft 
state region (see Fig.~\ref{gams}). Presumably such a bimodality is seen in the 
time lags, where for 2.1 $<$ $\Gamma_{1}$ $<$ 2.7 the lags are much larger than 
for the rest of the soft state $\Gamma_{1}$ range \citep{Grinberg2014}, 
possibly exposing the soft intermediate state. The SIMS state is characterized 
by a clearly smaller rms amplitude and this is seen in figs 3 and 4 of 
\citet{Grinberg2014}. A similar effect was already observed by 
\citet{Pottschmidt2003a} during failed transitions. At hard X-rays we found the 
opposite behavior: variability amplitude $S_{\rm V}$ for the HI state is clearly 
smaller than for the SI state (see Table~\ref{6mean}). The radio emission is 
rather weak for the SI state and very strong with flares for the HI state, in 
agreement with the SIMS and HIMS states characteristics \citep{Belloni2016}. 
Another similarity is that our HI state occupies quite a wide range of 
$\Gamma_{\rm H}$ as the HIMS does for the hardness ratio, whereas the SI and 
SIMS regions are quite narrow (see fig. 3.7 of \citealt{Belloni2010}).

The pure soft state term is used for the spectral state dominated in the soft 
X-ray band ($<$ 20 keV) by the disk emission \citep{Wilms2001}. Although the 
state selection adopted in \citet{Filothodoros2018} does not completely 
correspond to the six states found here, for a summed \integrale spectrum of the 
softest Cyg X-1 data they found the hard/soft compactness ratio, $l_{\rm 
h}/l_{\rm s}$, and optical depth of the plasma sharply smaller than the
corresponding values for the rest of soft state spectra. Thus, spectral analysis 
with a hybrid Comptonization model is consistent with a separation of the pure 
soft state through the $\Gamma_{\rm H}$-$F_{\rm H}$ diagram. 

\subsection{System geometry}
\label{geo}

Provided that the dominant process shaping the hard X-ray continuum of BHBs is 
inverse Comptonization on thermal electrons, a primary driver of the spectral 
slope in that band is the system geometry considered in terms of the Compton 
amplification strength \citep{Haardt1991,Stern1995}. This amplification can be 
either modeled directly through a certain parameter of a given implementation of 
the Comptonization or derived from fluxes of the seed photons and Comptonized 
component. It has been demonstrated many times that the photon index strongly 
correlates with the Compton parameter $y$, compactness ratio $l_{\rm h}/l_{\rm 
s}$ or the Compton amplification factor (e.g. 
\citealt{Malzac2001a,Wilms2006,Gierlinski2010}). Therefore, $\Gamma_{\rm H}$ 
parameter can be used as a tracer of the system geometry. 

The distribution of the $\Gamma_{\rm H}$ values found by us for the three harder 
states is quite similar to those found for Seyfert nuclei 
\citep{Lubinski2016a}. The PH and TH state peaks correspond to radio-quiet 
Seyferts, having the hardest X-ray spectra and a weak or moderate radio 
emission. On the other hand, several radio-loud objects of \citet{Lubinski2016a} 
sample exhibit softer spectra resembling the HI state of Cyg X-1. Both photon 
index distributions, that of Cyg X-1 and that of Seyferts, show an abrupt 
cut-off below $\Gamma$ = 1.7 (see fig. 3 of \citealt{Lubinski2016a}), i.e., just 
below the main peak. 

Interestingly enough, such a peak of $\Gamma$ around 1.7 can be justified on a 
theoretical ground. In their study of the synchrotron boiler effect 
\citet{Malzac2009} found that assuming an initially non-thermal distribution of 
plasma electrons in the absence of an accretion disk, a quasi-thermal 
Comptonization continuum can be obtained for a wide range of the plasma 
compactness, with the photon index close to 1.7. Also their range of $kT_{\rm 
e}$ values, concentrated between 30--50 keV, is similar to that observed for 
both hard state of Cyg X-1 \citep{Zdziarski2002b,DelSanto2013} and hard spectra' 
Seyferts \citep{Lubinski2016a}. A similar result with a stable spectral slope was 
obtained by \citet{Poutanen2009}, who also investigated the synchrotron boiler 
effect for an initially non-thermal plasma. Therefore, the pure hard state of 
Cyg X-1 can be interpreted as a limiting mode of accretion dominated by a hot 
compact plasma, with a negligible interaction with the accretion disk, probably 
truncated at large radii. The seed photons undergoing Comptonization are 
produced by a synchrotron radiation of the electrons of the plasma region itself 
(e.g. \citealt{Veledina2011,Poutanen2018}). A quite narrow distribution of 
$\Gamma_{\rm H}$ (or $y$), peaking at the same value for Cyg X-1 and Seyferts, 
can indicate that the system geometry is reaching its physical limit, with the 
plasma region being concentrated well within the inner radius of the disk.

The pure soft state can be interpreted as a second limiting geometry of the Cyg 
X-1 
accreting system, with no signature for an autonomous plasma region besides the 
non-thermal flares and atmosphere of the accretion disk. The PS state data 
dominate the softest (s1) spectrum analyzed by \citet{Filothodoros2018}. That 
spectrum, with the compactness ratio $l_{h}/l_{s}$ of 0.04$\pm0.01$, is softer 
than any \xtee spectrum analyzed by \citet{Wilms2006} and 
\citet{Gierlinski2010}, with all values of this parameter $>$0.2. On the other 
hand, there were extremely soft spectra of Cyg X-1 observed by \suzakue in 2010 
and 2013, contemporary to our PS state. The softest of these was fitted with 
$l_{h}/l_{s}$ = 0.11$_{-0.01}^{+0.02}$ \citep{Kawano2017}, however, a closer 
comparison is impossible, due to several parameters fixed at different values 
than in \citet{Filothodoros2018} and the \suzakue spectrum fitted in the 0.8--60 
keV band, whereas the \integrale spectra were fitted in the 3--200 keV band. The 
fact that the PH state was not observed with \xtee can be explained by much 
larger time spent by Cyg X-1 in the soft state in the 2010--2017 period than 
during the \xtee observations.

In summary, the Cyg X-1 system appears to evolve between the two extreme plasma 
states with (primarily for the TH state) non-thermal electrons, whereas for the 
four transient states the plasma is mostly thermal. States change presumably due 
to the changes of geometry of the plasma-disk system, resulting in a varying 
cooling of the plasma by the disk photons. The fact that this evolution goes 
through several distinct plasma states instead of some gradual transformation 
demands an explanation. Collecting new \integrale data in the coming years will shed 
more light on the location of these states in the $\Gamma_{\rm H}$-$F_{\rm H}$ 
diagram, especially those less populated. Among the four transient states the 
most intriguing appears the soft intermediate state, showing vertical 
orientation in Fig.~\ref{gfcon}. This vertical structure has a width comparable 
with the typical uncertainty of $\Gamma_{\rm H}$ in this range of the diagram. 
Thus, the intrinsic scatter of the photon index is presumably quite small, 
indicating a well defined geometry of the hard-to-soft regime transition, a kind 
of some "bottleneck" mechanism.

Since clustering is observed for both hard and soft regimes, its origin might be 
a certain common mechanism forming different plasma geometries. A varying 
accretion rate is thought to be a principal driver of the state transition in 
binary systems. Nevertheless, even if the system accretes at distinct rates, 
these rates should be well preserved during the mass transport in the disk. In 
addition, there should be a tight mechanism of forming specific plasma 
geometries in a reaction to a varying accretion rate. More light on the plasma 
states in Cyg X-1 will be shed after state-wise broad-band spectral analysis 
with physical models and more advanced timing analysis.

\section{Conclusions}

We have analyzed \integral/ISGRI data collected over 15 years of Cyg X-1 
observations, exploring all uninterrupted monitoring periods, lasting typically 
0.5--2 hour. This data set comprised almost 8000 pointings for which we have 
extracted spectra and computed the fractional variability amplitude. Since the 
emission of Cyg X-1 in the 22--100 keV is unabsorbed and weakly affected by the 
Compton reflection and high-energy thermal cut-off, even a simple power-law 
model allows for a reliable characterization of the primary emission from the 
hot plasma. Using this model we determined the flux $F_{\rm H}$ and photon index 
$\Gamma_{\rm H}$ for the 22--100 keV band. To explore the hard X-ray/radio 
relation we used the RT/AMI data at 15 GHz. Our main findings are as follows: \\
1. The $\Gamma_{\rm H}$-$F_{\rm H}$ density diagram reveals six distinct 
regions, concentrated around $\Gamma_{\rm H}$ = 1.7, 1.85, 2.0, 2.2, 2.5 and 
2.8, with a relative population of 33\%, 21\%, 19\%, 6\%, 14\% and 7\%, 
respectively. These six plasma states of Cyg X-1 were named, accordingly, pure 
hard, transitional hard, hard intermediate, soft intermediate, transitional 
soft and pure soft state. Such clustering is observed for the first time for any 
BH binary. \\
2. Each of the six plasma states exhibits a different range of variability measured 
with the fractional variability amplitude $S_{\rm V}$. In the three softer 
states the mean $S_{\rm V}$ is typically $\gtrsim$ 10\%, whereas for the three harder 
states it is typically $\lesssim$ 5\%, reaching minimal values for the 
transitional hard state. Our results extend to higher energy the findings of 
\citet{Grinberg2014}, who found for Cyg X-1 a minimal variability for medium 
hardness data at several energy bands below 15 keV. This trend is different from 
the typical behavior observed in soft X-rays for transient BHBs, where the 
variability increases monotonically with the spectral hardness. \\ 
3. The radio flux $F_{\rm R}$ at 15 GHz is correlated with all three hard X-ray 
observables. An overall correlation with the maximal radio fluxes seen for an 
intermediate range of hard X-ray flux and spectral slope values, was already
reported for Cyg X-1. However, the radio flux plotted against the hard photon 
index occupies different levels for all six states. In addition, we found that 
the radio flux decreases with an increasing hard X-ray  variability amplitude 
$S_{\rm V}$. \\
4. We have confronted our plasma states with the standard spectral states 
selected using the softer X-ray bands. We identify the pure hard and transitional
hard states with 
the two hard states found by \citet{Pottschmidt2003a} and \citet{Grinberg2014}, 
both using the soft X-rays. A distinction between the two intermediate states 
for Cyg X-1 is made for the first time, they appear to be counterparts of the 
HIMS and SIMS states observed typically for BHB transients. The pure soft state 
was identified within the soft state through its specific hard X-ray
variability. \\
5. Under the assumption of the Comptonization process as a primary source of the 
hard X-ray radiation our results can be interpreted in terms of six distinct 
geometries of the plasma region in Cyg X-1. The hardest and softest states show 
no hour-scale $F_{\rm H}$-$\Gamma_{\rm H}$ correlation, suggesting a lack of 
plasma-accretion disk interaction. The other four states exhibit strong 
correlation, slowly decreasing with increasing $\Gamma_{\rm H}$, presumably due 
to decreasing inner radius of the disk.\\
6. Our results for the pure hard state agree with the predictions of the synchrotron 
boiler models, with primarily non-thermal hot electrons Comptonizing their own 
synchrotron radiation. Such models predict also a narrow distribution of the 
photon index for a broad range of the system parameters, in agreement with our 
findings.\\ 

\section*{Acknowledgments}
This research was based on observations with \integral, an ESA project with 
instruments and science data centre funded by ESA member states (especially the 
PI countries: Denmark, France, Germany, Italy, Switzerland, Spain), the Czech 
Republic, and Poland and with the participation of Russia and the USA. We have
been supported by the Polish National Science Centre (NCN) grant 
2014/13/B/ST9/00570 (AF,PL) and grant 2015/18/A/ST9/00746 (AAZ). This research 
has made use of MAXI data provided by RIKEN, JAXA and the MAXI team. We thank 
the reviewer for his/her thorough review and highly appreciate the comments and 
suggestions, which significantly contributed to improving the quality of the 
publication.

\vspace{5mm}

\facilities{INTEGRAL(IBIS,JEM-X,SPI),AMI,\\
RXTE(ASM),MAXI,Swift(BAT),CGRO(BATSE).}

\software{OSA v10.2 and 11.0 (Courvoisier et al. 2003), mxproduct 
(Matsuoka et al. 2009), HEAsoft (HEASARC 2014), XSPEC (v12.9.0; Arnaud 1996)}

\appendix

\section{Separation of the two soft states}
\label{appena}

\begin{figure*}[h!]
\begin{center}
\includegraphics[width=8.5cm]{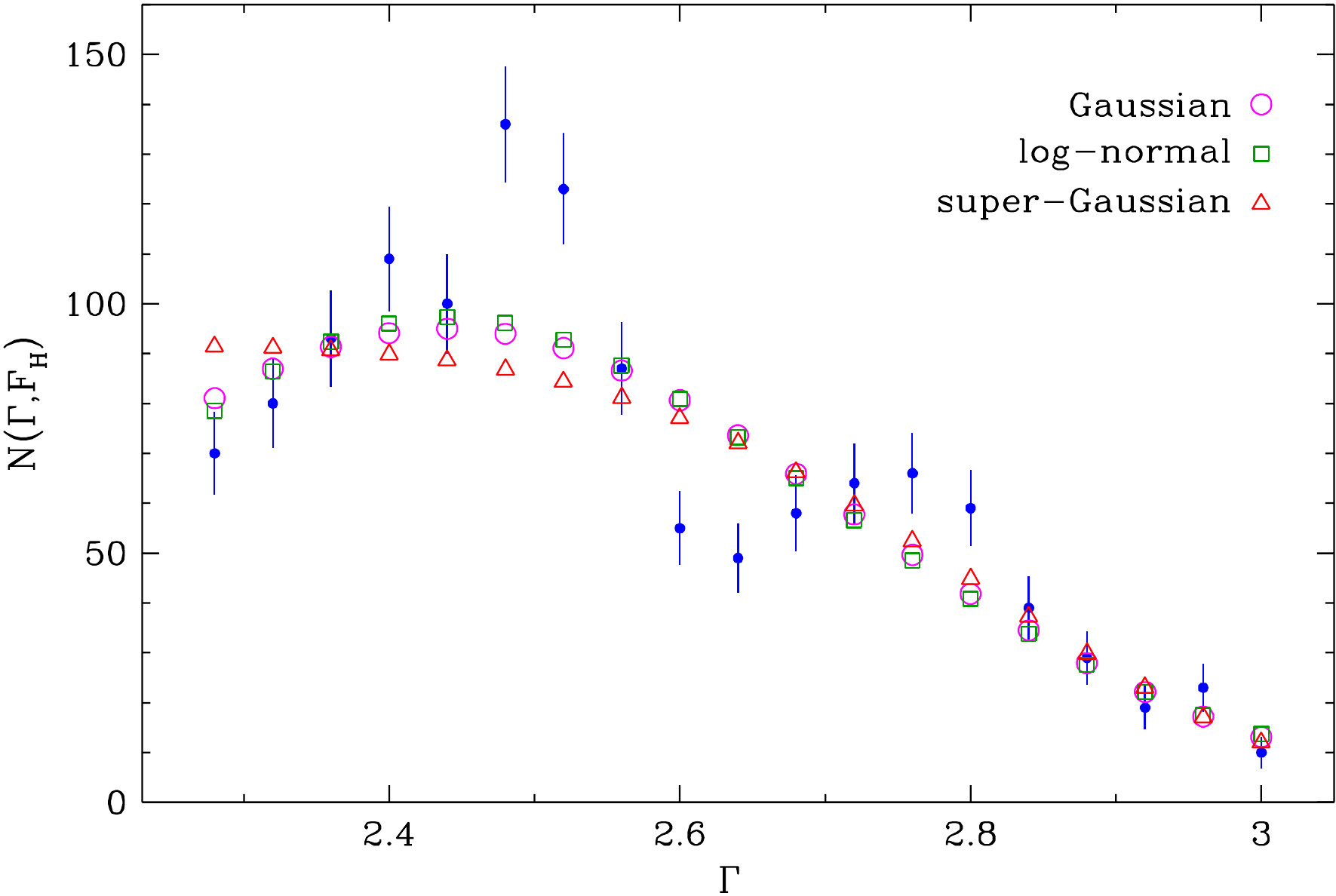}
\end{center}
\caption{Projection of the $\Gamma_{\rm H} - F_{\rm H}$ diagram data on the 
$\Gamma_{\rm H}$ axis. Number of points within a rectangular region centered at 
given $\Gamma_{\rm H}$ is shown in blue. Fitted distributions are shown with 
symbols without error bars.}
\label{ridge}
\end{figure*}

The number of $\Gamma_{\rm}$-$F_{\rm H}$ data missing in the valley between the 
transitional soft and pure soft states was determined using rectangular regions 
with the longer side perpendicular to the line connecting the centers of these 
two regions (see Sec.~\ref{plasma} and Fig.~\ref{gfcon}). Figure~\ref{ridge} 
presents the number of data points found for rectangular regions centered at 
$\Gamma_{\rm H}$ between 2.28 and 3.0, with a step of 0.04. The null hypothesis 
probability (i.e., a single soft state) was tested with the three density 
models: Gaussian, log-normal, and super-Gaussian fitted to the rectangular regions 
data. The super-Gaussian is a flat-top Gaussian, where the standard Gaussian 
exponent argument is raised additionally to some power, in our case it was set 
to 2: $f(x) \propto \exp \{{-[(x-x_{0})^{2}/(2\sigma^{2})]^{2}}\}$. The 
$\chi^{2}$ test values for the log-normal, Gaussian and super-Gaussian models 
are 62.4, 64,5, and 73.4, respectively, corresponding to $p$-values of 2$\times 
10^{-7}$, 9$\times 10^{-8}$, and $< 10^{-8}$, respectively, for 16 degrees of 
freedom. The valley between the TS and PS states is seen at $\Gamma_{\rm H}$ 
around 2.6. For all tested models at least 60 data points are missing in the 
valley and must be shifted to either the transitional soft ($\Gamma_{\rm H} 
\approx$ 2.5) or to the pure soft state ($\Gamma_{\rm H} \approx$ 2.78) peaks. 

\section{Clustering test with the eqpair model}
\label{appenb}

To simulate a set of realistic spectra we used the results of 
\citet{Filothodoros2018} who fitted the hybrid Comptonization model 
\texttt{eqpair} to the JEM-X and ISGRI spectra of Cyg X-1 in the 3--200 keV
band. Their spectra were grouped according to the 40--100/22-40 keV hardness 
ratio, forming 12 spectral sets, 6 for hard state and 6 for soft state (see 
fig.~1 in \citet{Filothodoros2018}). Since \citet{Filothodoros2018} and our data 
cover the same spectral shape range, their hardness ratio range corresponds to 
our $\Gamma_{\rm H}$ range. The leading \texttt{eqpair} parameter driving the 
spectral slope is the hard/soft compactness ratio, $l_{h}/l_{s}$. Its dependence 
on the 40--100/22-40 keV hardness found for Cyg X-1 by \citet{Filothodoros2018} 
is nearly exponential, with $\log(l_{h}/l_{s})$ between -0.9 and 1.3. Using 
their results we found that the dependence of the five other main \texttt{eqpair} 
parameters on $\log(l_{h}/l_{s})$ is practically linear, separately for the 
hard and soft state. Values of these parameters were computed for 170 
$\log(l_{h}/l_{s})$ channels in the [0.45,1.3] range for the hard state, and for 
200 $\log(l_{h}/l_{s})$ channels in the [-0.9,0.1] range for the soft state. These 
five parameters are (with the limits found for the hard and soft state, 
respectively, given in parenthesis): Thomson scattering depth of the plasma 
(1.52--0.88, 0.7--0.12), non-thermal/thermal power ratio (0.78--0.51, 
0.57--0.31), index of the non-thermal electrons' power-law distribution 
(2.04--2.25, 2.13--3.74), Compton reflection strength (0.14--0.6, 1.81--3.42) 
and the seed photons temperature (150 eV, 150--300 eV). For each simulation we 
have randomly chosen a spectrum from our 7821 spectra (see Sec.~\ref{inteda}), 
using its exposure time and response files to produce a fake spectrum. In total 
we simulated 300,000 spectra for each of the hard and soft state regimes. The 
22--100 keV flux for each channel was randomized with a normal distribution,
approximating that observed in Fig.~\ref{gfcon}. 

\begin{figure*}[h!]
\begin{center}
\includegraphics[width=7.5cm,angle=270]{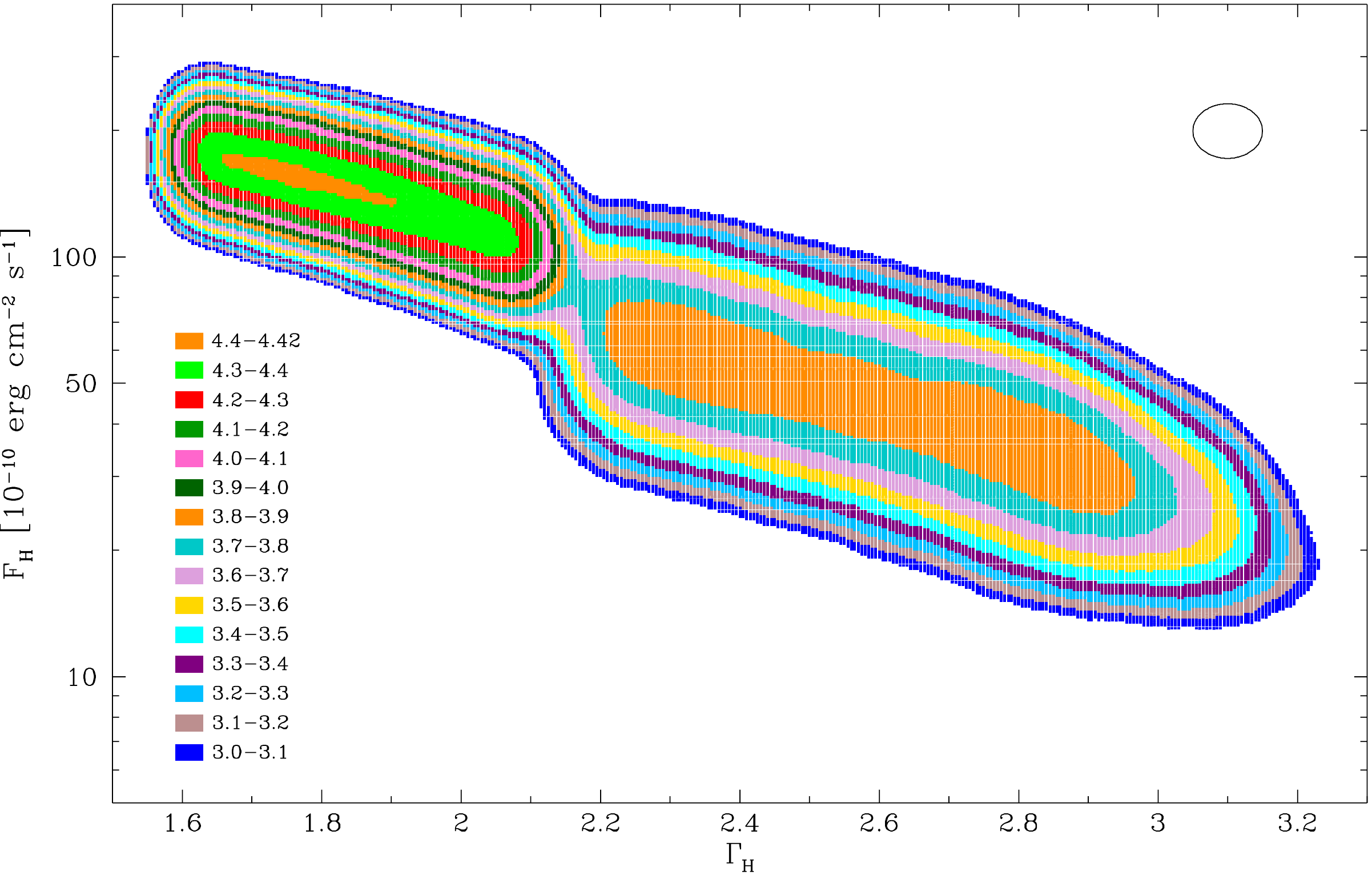}
\end{center}
\caption{Density map of the $\Gamma_{\rm H}$-$F_{\rm H}$ distribution obtained 
from the power-law model fitted to the spectra simulated with the 
\texttt{eqpair} model. The ellipse shows the size of the density sampling region.}
\label{gfsim}
\end{figure*}

The simulated spectra were then fitted with the power-law model, producing the set
of the $\Gamma_{\rm H}$-$F_{\rm H}$ results. The corresponding density map is shown
in Fig.~\ref{gfsim}, with a smooth distribution of density for both the hard and
soft regimes. We tested various modifications of the functions approximating the 
dependence of the \texttt{eqpair} parameters on $\log(l_{h}/l_{s})$. Only small
($<$ 10\%) and broad density maxima are occasionally observed. We did not 
obtain an artificial aggregation of the photon index resembling the six plasma 
states shown in Fig.~\ref{gfcon}.

\bibliography{plnew}

\end{document}